\documentclass[pre,aps,twocolumn]{revtex4}
\usepackage{graphicx}
\usepackage{dcolumn}
\usepackage{bm}
\usepackage{amsmath,amssymb}
\usepackage{color,soul}
\usepackage[toc,page]{appendix}

\begin{document}

\title{Unconventional strengthening of a bipartite entanglement of a mixed spin-(1/2,1) Heisenberg dimer achieved through Zeeman splitting}
\author{Hana \v Cen\v carikov\'a$^{1,\footnote{Corresponding author: hcencar@saske.sk}}$ and Jozef Stre\v{c}ka$^2$}
\affiliation{
$^1$Institute of Experimental Physics, Slovak Academy of Sciences, Watsonova 47, 040 01 Ko\v {s}ice, Slovakia \\ 
$^2$ Department of Theoretical Physics and Astrophysics, Faculty of Science, P. J. \v{S}af\'{a}rik University, Park 
Angelinum 9, 040 01 Ko\v{s}ice, Slovakia}

\begin{abstract}
The bipartite quantum and thermal entanglement is quantified within pure and mixed states of a mixed spin-(1/2,1) Heisenberg dimer with the help of negativity. It is shown that the negativity, which may serve as a measure of the bipartite entanglement at zero as well as nonzero temperatures, strongly depends on intrinsic parameters as for instance exchange and uniaxial single-ion anisotropy in addition to extrinsic parameters such as temperature and magnetic field. It turns out that a rising magnetic field unexpectedly  reinforces the bipartite entanglement due to the Zeeman splitting of energy levels, which lifts a two-fold degeneracy of the quantum ferrimagnetic ground state. The maximal bipartite entanglement is thus reached within a quantum ferrimagnetic phase at sufficiently low but nonzero magnetic fields on assumption that the gyromagnetic g-factors of the spin-1/2 and spin-1 magnetic ions are equal and the uniaxial single-ion anisotropy is a half of the exchange constant. It is suggested that the heterodinuclear complex [Ni(dpt)(H$_2$O)Cu(pba)]$\cdot$2H$_2$O (pba=1,3-propylenebis(oxamato) and dpt=bis-(3-aminopropyl)amine), which affords an experimental realization of the mixed spin-(1/2,1) Heisenberg dimer, remains strongly entangled up to relatively high temperatures (about 140~K) and magnetic fields (about 140~T) being comparable with the relevant exchange constant. 
\end{abstract}

\maketitle

\section{Introduction}
\label{sec:introduction}

Molecular magnetic materials \cite{jong74,carl86,kahn93,siek17} belong to prominent solid-state resources for quantum computation and quantum information processing \cite{leue01}, because they are capable of building extremely dense and efficient memory devices implementing Grover's algorithm \cite{grov97}. The Grover's search algorithm requires a superposition of 'single-particle' quantum states, whereas spin states of a single magnetic molecule with a sufficiently long relaxation time provide eligible platform for its technological implementation on the grounds of single-molecule magnets \cite{thom96,sang97,gatt06}. Some quantum algorithms, such as Shor's factoring algorithm \cite{shor94}, however require both superposition and entanglement of 'many-particle' quantum states, which naturally occur in many-particle quantum spin systems forming basic building blocks of molecular-based magnetic materials. 

Many-particle quantum spin systems have been extensively investigated in the context of quantum information processing due to a possibility of creation and distribution of quantum entanglement between specific spin units acting as qubits \cite{amic08} as well as the speed-up in quantum computation and communication \cite{arne01}. It is noteworthy that entanglement measures such as negativity \cite{pere96,horodecki96,vidal02} or concurrence \cite{woot98} can be related via certain witnesses to thermodynamic quantities \cite{waza02,aldo08}, which additionally offer an intriguing possibility for experimental testing \cite{souz08,cruz16}. Some quantum protocols such as a quantum teleportation of information cannot be even realized without many-particle entangled states \cite{yeoy02,rojm17,frei19,zhen19}.  

Bearing all this in mind, it appears worthwhile to investigate how a degree of entanglement in quantum spin systems is affected by extrinsic parameters such as temperature and external magnetic field. From the viewpoint of possible technological applications, it is especially important to find out whether a quantum entanglement emergent at absolute zero temperature may persists as a thermal entanglement at sufficiently high temperatures. The quantity concurrence has been widely used in order to capture a strength of the bipartite thermal entanglement in several spin-1/2 quantum systems: dimer \cite{arne01,wang01}, trimer \cite{bose05}, tetramer \cite{bose05,karl20}, chain \cite{wang02}, ladder \cite{trib09,roja16}, tube \cite{stre16,alec16}, tetrahedral chain \cite{roja13,stre14}, trimerized chain \cite{stre17}, diamond chain \cite{torr14,qiao15,roja17,zhen18,carv18,carv19}, pentagonal chain \cite{karl18} and branched chain \cite{karl19,souz19}. On the other hand, one may rely on the concept of negativity in order to capture a bipartite thermal entanglement of the mixed spin-1/2 and spin-1 quantum systems, which were however much less comprehensively investigated in comparison with their single spin-1/2 counterparts \cite{sunz06,sunz09,suny09,sola11,soln11,abga15}. The numerous previous studies serve in evidence that the thermal entanglement does not have to be necessarily suppressed upon increasing of temperature and magnetic field in opposite to naive expectations. 

In the present paper, we will examine a strength of the quantum and thermal entanglement within pure and mixed states of a mixed spin-(1/2,1) Heisenberg dimer with the exchange and uniaxial single-ion anisotropy in presence of the external magnetic field. The present theoretical study is motivated by the molecular-based compound [Ni(dpt)(H$_2$O)Cu(pba)]$\cdot$2H$_2$O (pba=1,3-propylenebis(oxamato) and dpt=bis-(3-aminopropyl)amine) \cite{hagi99}, which could be classified as a heterodinuclear complex of the exchange-coupled spin-1/2 Cu$^{2+}$ and spin-1 Ni$^{2+}$ magnetic ions to be further abbreviated as the CuNi compound.

The organization of this paper is as follows. An exact calculation for the negativity of the mixed spin-(1/2,1) Heisenberg dimer in a magnetic field is presented in Sec. \ref{sec:model}. The most interesting results for the quantum and thermal entanglement of the mixed spin-(1/2,1) Heisenberg dimer will be presented in Sec. \ref{sec:result} as function of the exchange anisotropy, the uniaxial single-ion anisotropy and magnetic field together with the relevant theoretical prediction for the CuNi complex. A brief summary of the most important scientific achievements is presented in Sec. \ref{sec:conc} along with future outlooks and perspectives.

\section{Model and Method}
\label{sec:model}
In the present paper, we will investigate in detail a quantum and thermal entanglement of the mixed spin-(1/2,1) Heisenberg dimer defined by the Hamiltonian
\begin{eqnarray}
\hat{\cal H}\!\!\!&=&\!\!\!J\left[\Delta(\hat{S}^x\hat{\mu}^x\!+\!\hat{S}^y\hat{\mu}^y)\!+\!\hat{S}^z\hat{\mu}^z\right]\!+\!D(\hat{\mu}^z)^2
\nonumber\\
\!\!\!&-&\!\!\!
g_1\mu_BB\hat{S}^z\!-\!g_2\mu_B B\hat{\mu}^z,
\label{eq1}
\end{eqnarray} 
where $\hat{S}^{\alpha}$ and $\hat{\mu}^{\alpha}$ ($\alpha\!=\!x,y,z$) denote spatial components of the spin-1/2 and spin-1 operators, respectively. The coupling constant $J$ determines the Heisenberg exchange interaction between the spin-1/2 and spin-1 magnetic ions, the parameter $\Delta$ determines the XXZ exchange anisotropy in this exchange interaction and the parameter $D$ is an uniaxial single-ion anisotropy acting on the spin-1 magnetic ions only. Finally, the parameter $B$ denotes a static external magnetic field, $\mu_B$ is Bohr magneton, while $g_1$ and $g_2$ are Land\'e $g$-factors of the spin-1/2 and spin-1 magnetic ions, respectively. 

A matrix representation of the Hamiltonian (\ref{eq1}) in the standard basis formed by the eigenvectors $\left\vert \varphi_i\right\rangle\in \left\{ \left\vert\frac{1}{2},1\right\rangle, \left\vert\frac{1}{2},0\right\rangle, \left\vert\frac{1}{2},-1\right\rangle, \left\vert-\frac{1}{2},1\right\rangle, \left\vert-\frac{1}{2},0\right\rangle, \left\vert-\frac{1}{2},-1\right\rangle \right\}$  of $z$-components of the constituting spin-1/2 and spin-1 entities reads as follows
\begin{align}
&\left\langle \varphi_j \right\vert \hat{\cal H} \left\vert\varphi_i \right\rangle
=\left(
\begin{array}{cccccc}
 H_{11}& 0 & 0& 0 & 0 & 0\\
0 &H_{22} & 0 &H_{24}& 0 & 0 \\
0 & 0 & H_{33} & 0 & H_{35} & 0\\
0 & H_{42} & 0 & H_{44} & 0 & 0\\
0 & 0 & H_{53}& 0 & H_{55} & 0  \\
0 & 0& 0 & 0 & 0 & H_{66}
\end{array}\right),
\label{eq1a}
\end{align}
whereas six diagonal elements are defined by 
\begin{align}
&H_{11}=\frac{1}{2}\left[J\!+\!2D\!-\!(h_1\!+\!2h_2)\right],\nonumber\\
&H_{22}=-\frac{h_1}{2},\nonumber\\
&H_{33}=-\frac{1}{2}\left[ J\!-\!2D\!+\!(h_1\!-\!2h_2)\right],\nonumber\\
&H_{44}=-\frac{1}{2}\left[ J\!-\!2D\!-\!(h_1\!-\!2h_2)\right],\nonumber\\
&H_{55}=\frac{h_1}{2}\;,\nonumber\\
&H_{66}=\frac{1}{2}\left[J\!+\!2D\!+\!(h_1\!+\!2h_2)\right],\label{eq1b}
\end{align}
and four off-diagonal elements are equal to 
\begin{align}
H_{24}=H_{42}=H_{35}=H_{53}=\frac{J\Delta}{\sqrt{2}}.
\label{eq1c}
\end{align}
For abbreviation purposes, we have introduced in above two new parameters $h_{1}\!=\!g_{1}\mu_BB$ and $h_{2}\!=\!g_{2}\mu_BB$ related to 'local' Zeeman terms (magnetic fields) acting on the spin-1/2 and spin-1 magnetic particles, which may be different due to difference of the gyromagnetic g-factors $g_1 \neq g_2$. A relatively simple (sparse) structure of the Hamiltonian matrix (\ref{eq1a}) allows us to obtain a complete set of eigenvalues by an exact analytical diagonalization 
\begin{align}
E_{1,2}&=\frac{1}{2}\left[J\!+\!2D\!\mp\!(h_1\!+\!2h_2)\right],
\label{eq2}\\
E_{3,4}&=-\frac{1}{4}\left(J\!-\!2D\!+\!2h_2\right)
\nonumber\\
&\mp\frac{1}{4}
\sqrt{\left[J\!-\!2D\!-\!2(h_1\!-\!h_2)\right]^2\!+\!8(J\Delta)^2},
\label{eq3}\\
E_{5,6}&=-\frac{1}{4}\left(J\!-\!2D\!-\!2h_2\right)
\nonumber\\
&\mp\frac{1}{4}
\sqrt{\left[J\!-\!2D\!+\!2(h_1\!-\!h_2)\right]^2\!+\!8(J\Delta)^2},
\label{eq4}
\end{align} 
whereas the corresponding eigenvectors read
\begin{align}
\displaystyle
&\left\vert\psi_1\right\rangle=\left\vert\frac{1}{2},1\right\rangle, 
\label{eq5a}\\
&\left\vert\psi_2\right\rangle=\left\vert-\frac{1}{2},-1\right\rangle, 
\label{eq5}\\
&\left\vert\psi_{3,4}\right\rangle=c_1^{\mp}\left\vert\frac{1}{2},0\right\rangle \mp c_1^{\pm}\left\vert-\frac{1}{2},1\right\rangle,
\label{eq6}\\
&\left\vert\psi_{5,6}\right\rangle=c_2^{\pm}\left\vert\frac{1}{2},-1\right\rangle \mp c_2^{\mp}\left\vert-\frac{1}{2},0\right\rangle.
\label{eq7}
\end{align} 
The last four eigenvectors \eqref{eq6} and \eqref{eq7} are defined through the probability amplitudes
\begin{align}
c_1^{\pm}&=\frac{1}{\sqrt{2}}\sqrt{1\!\pm\!\frac{J\!-\!2D\!-\!2(h_1\!-\!h_2)}{\sqrt{\left[J\!-\!2D\!-\!2(h_1\!-\!h_2)\right]^2\!+\!8(J\Delta)^2}}
}\;, 
\nonumber\\
c_2^{\pm}&=\frac{1}{\sqrt{2}}\sqrt{1\!\pm\!\frac{J\!-\!2D\!+\!2(h_1\!-\!h_2)}{\sqrt{\left[J\!-\!2D\!+\!2(h_1\!-\!h_2)\right]^2\!+\!8(J\Delta)^2}}
}\;.
\label{eq8}
\end{align} 
\\
To explore a degree of quantum and thermal entanglement in pure and mixed states of the mixed spin-(1/2,1) Heisenberg dimer one may employ the quantity negativity~\cite{pere96,horodecki96,vidal02}
\begin{eqnarray}
{\cal N}\!=\!\sum_{i=1}^{6}\frac{|\lambda_i|-\lambda_i}{2}\;,
\label{eq9}
\end{eqnarray} 
which is defined through eigenvalues $\lambda_i$ of a partially transposed density matrix $\rho^{T_{1/2}}$ derived from the overall density matrix $\rho$ upon a partial transposition $T_{1/2}$ with respect to one subsystem. In this particular case $T_{1/2}$ denotes a partial transposition with respect to states of the spin-1/2 magnetic ion. According to separability criterion invented for partially transposed density matrices by Peres~\cite{pere96} the negativity becomes zero (${\cal N}=0$) for separable (factorizable) states, while it becomes non-zero (${\cal N}\neq0$) for entangled (nonseparable) states. Consequently, the necessary prerequisite for detecting a quantum or thermal entanglement within pure or mixed states of the mixed spin-(1/2,1) Heisenberg dimer is at least one negative eigenvalue $\lambda_i$ of the partially transposed density matrix $\rho^{T_{1/2}}$.

The density operator $\hat{\rho}$ of the mixed spin-(1/2,1) Heisenberg dimer can be easily calculated from the eigenvalues (\ref{eq2})-(\ref{eq4}) and the respective eigenvectors (\ref{eq5a})-(\ref{eq7}) according to the formula
\begin{align}
\hat{\rho}&=\frac{1}{Z}\sum_{i=1}^6 \exp(-\beta E_i) \left\vert\psi_i\right\rangle \left\langle \psi_i\right\vert,
\label{eq10}
\end{align} 
which is expressed in terms of the partition function $Z = \sum_{i=1}^6 \exp(-\beta E_i)$ acquiring the following explicit form
\begin{align}
Z&=2\left\{
{\rm e}^{-\frac{\beta}{2} (J+2D)}\cosh\left[\frac{\beta}{2} (h_1\!+\!2h_2) \right]\!+\! {\rm e}^{\frac{\beta}{4}(J-2D)} \times \right.
\nonumber\\
&\left[{\rm e}^{\frac{\beta h_2}{2}} \cosh\left(\frac{\beta}{4}\sqrt{\left[J\!-\!2D\!-\!2(h_1\!-\!h_2)\right]^2\!+\!8(J\Delta)^2}\right) \right.
\nonumber\\
&+\left.\left.{\rm e}^{-\frac{\beta h_2}{2}}\cosh\left(\frac{\beta}{4}\sqrt{\left[J\!-\!2D\!+\!2(h_1\!-\!h_2)\right]^2\!+\!8(J\Delta)^2}\right)
\right]\right\}\!.
\label{eq11}
\end{align} 
Of course, the density matrix corresponding to the density operator \eqref{eq10} has similar matrix structure as the Hamiltonian matrix (\ref{eq1a}) 
\begin{eqnarray}
{\rho}\!&=&\!\left(
\begin{array}{cccccc}
\rho_{11} & 0 & 0 & 0 & 0 & 0\\
0 & \rho_{22} & 0 & \rho_{24} & 0 & 0 \\
0 & 0 & \rho_{33} & 0 & \rho_{35} & 0\\
0 & \rho_{42} & 0 & \rho_{44} & 0 & 0\\
0 & 0 & \rho_{53}  &0 & \rho_{55} & 0\\
0 & 0 & 0 & 0 & 0 & \rho_{66} 
\end{array}
\right),
\label{eq12}
\end{eqnarray} 
whereas individual elements $\rho_{ij}$ of the density matrix are for the sake of brevity explicitly given in Appendix~\ref{App A}. A partial transposition $T_{1/2}$ with respect to states of the spin-1/2 magnetic ion gives the following partially transposed density matrix
\begin{eqnarray}
{\rho}^{T_{1/2}}\!\!\!&=&\!\!\!\left(
\begin{array}{cccccc}
\rho_{11} & 0 & 0 & 0 & \rho_{24} & 0\\
0 & \rho_{22} & 0 & 0 & 0 & \rho_{35}\\
0 & 0 & \rho_{33} & 0  & 0& 0\\
0 & 0 & 0 & \rho_{44} & 0 & 0\\
\rho_{24} & 0 &  0 & 0 & \rho_{55} & 0\\
0 & \rho_{35} & 0 & 0 & 0 & \rho_{66} 
\end{array}
\right).
\label{eq13}
\end{eqnarray} 
which has the following spectrum of eigenvalues
\begin{eqnarray}
\lambda_1\!\!\!&=&\!\!\! \rho_{33}, \\
\lambda_2\!\!\!&=&\!\!\! \rho_{44},\\
\lambda_{3,4}\!\!\!&=&\!\!\! \frac{\rho_{22}+\rho_{66}}{2}\!\pm\!\frac{1}{2}\sqrt{(\rho_{22}-\rho_{66})^2+4\rho_{35}^2}, \\
\lambda_{5,6}\!\!\!&=&\!\!\! \frac{\rho_{11}+\rho_{55}}{2}\!\pm\!\frac{1}{2}\sqrt{(\rho_{55}-\rho_{11})^2+4\rho_{24}^2}.
\label{eq14}
\end{eqnarray} 
It is quite clear that the eigenvalues $\lambda_4$ and $\lambda_6$ with negative sign before a square root may become, under certain conditions, negative, which is according to the definition \eqref{eq9} necessary prerequisite of nonzero negativity (bipartite entanglement). In the following part we will investigate in detail manifestation of a quantum and thermal entanglement of the mixed spin-(1/2,1) Heisenberg dimer depending on temperature, magnetic field, and magnetic anisotropy.

\section{Results and discussion}
\label{sec:result}

It is worthwhile to recall that suitable experimental realizations of the mixed spin-(1/2,1) Heisenberg dimer are offered by heterobimetallic complexes such as the CuNi compound \cite{hagi99} being composed of the exchange coupled spin-1/2 Cu$^{2+}$ and spin-1 Ni$^{2+}$ magnetic ions. Note furthermore that the transition-metal ions, as for instance Cu$^{2+}$ and Ni$^{2+}$, usually have gyromagnetic g-factors quite close to the spin-only value $g=2$ due to an almost full quenching of their orbital momentum \cite{jong74,carl86,kahn93}. In this regard, we will consider hereafter three different combinations of Land\'e g-factors: (i) $g_1=g_2=2.0$,  (ii) $g_1=2.2, g_2=2.0$, and (iii) $g_1=2.0, g_2=2.2$. The first case bears a close relation to the ideal case with equal gyromagnetic factors with the spin-only value $g=2$, while in the second and third case the gyromagnetic factor of the spin-1/2 magnetic ion Cu$^{2+}$ slightly exceeds that one of the spin-1 magnetic ion Ni$^{2+}$ or vice versa. For simplicity, the size of the antiferromagnetic exchange interaction $J>0$ may serve as an energy unit when defining a set of dimensionless quantities measuring a relative strength of the uniaxial single-ion anisotropy $D/J$, magnetic field $\mu_B B/J$ and temperature $k_B T/J$.  

\subsection{Quantum entanglement}
\label{ssec:gs}

\begin{figure}[t!]
{\includegraphics[width=.49\textwidth,trim=3cm 9.5cm 3.5cm 9.5cm, clip]{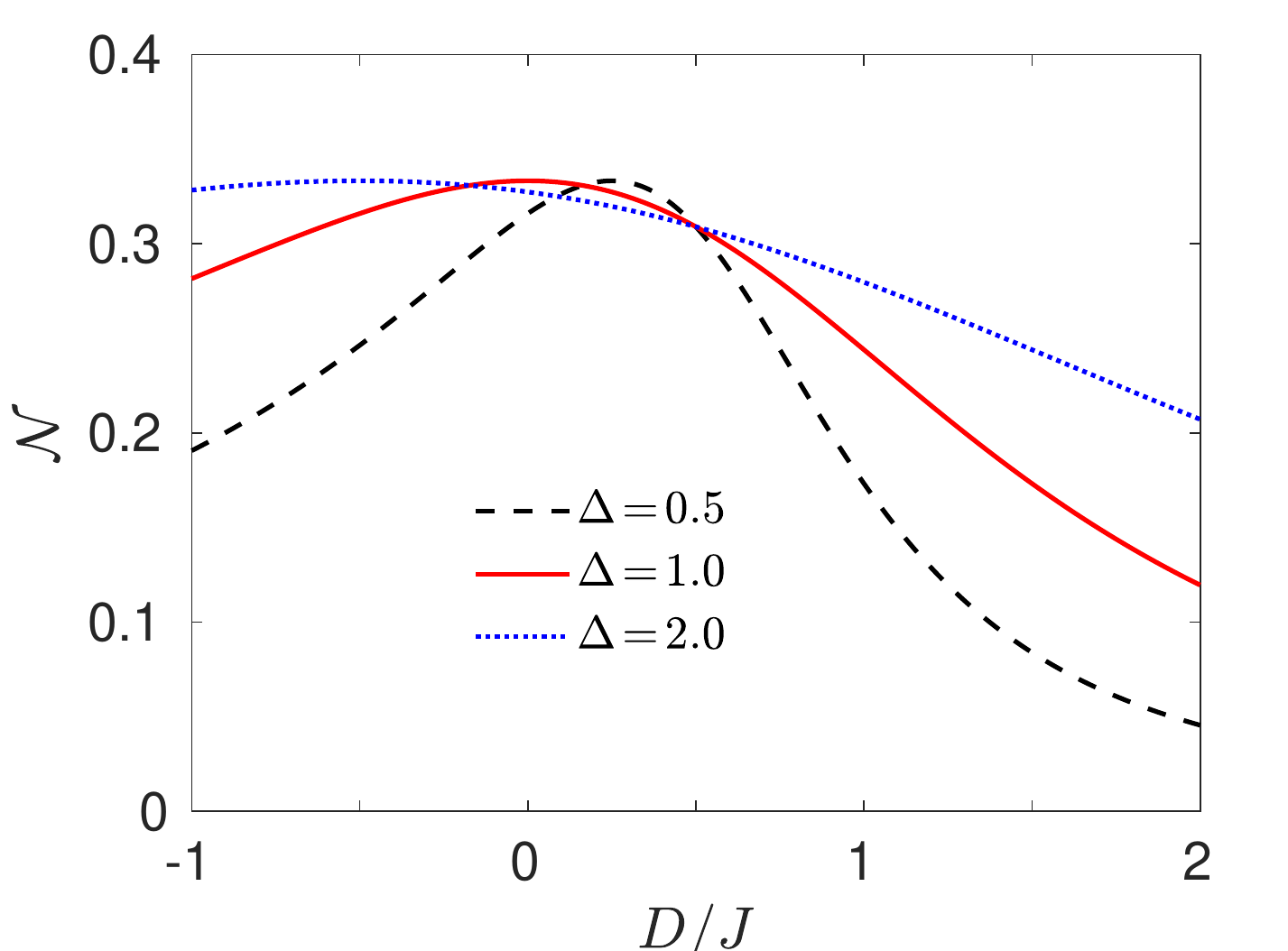}}
\caption{The negativity as a function of the uniaxial single-ion anisotropy $D/J$ for three different values of the exchange anisotropy $\Delta=0.5$, $1.0$, and $2.0$ at zero magnetic field.}
\label{fig1}
\end{figure}

First, our attention will be paid to a comprehensive analysis of a bipartite quantum entanglement of the mixed spin-(1/2,1) Heisenberg dimer at zero temperature and magnetic field depending on two intrinsic model parameters $\Delta$ and $D/J$ determining the exchange and uniaxial single-ion anisotropy, respectively. The zero-field ground state of the mixed spin-(1/2,1) Heisenberg dimer could be classified as a two-fold degenerate quantum ferrimagnetic state given by the eigenvectors
\begin{eqnarray}
| {\rm QFI}_{\pm} \rangle = \left\{ \begin{array}{ll} 
c_0^{+} | \!-\!1/2, 1 \rangle - c_0^{-} | 1/2, 0 \rangle, \\
c_0^{+} | 1/2, -1 \rangle - c_0^{-} | \!-\!1/2, 0 \rangle,
         \end{array} \right. 
\label{gs0}
\end{eqnarray}
where the probability amplitudes $c^{\pm}_0$ unambiguously determining the relevant quantum superposition of the microstates $| \!\mp\!1/2, 1 \rangle$ and 
$| \!\pm\! 1/2, 0 \rangle$ are given by 
\begin{eqnarray}
c_0^{\pm} = \frac{1}{\sqrt{2}} \left(\sqrt{1 \pm \frac{1-2\frac{D}{J}}{\sqrt{(1-2\frac{D}{J})^2 + 8\Delta^2}}} \right).
\label{pags0}
\end{eqnarray}
Using the respective density operator $\hat{\rho} = (\left\vert {\rm QFI}_{+} \right\rangle \left\langle {\rm QFI}_{+}\right\vert + \left\vert {\rm QFI}_{-} \right\rangle \left\langle {\rm QFI}_{-}\right\vert)/2$ one gets the following zero-temperature value of the negativity that characterizes the bipartite entanglement within a two-fold degenerate quantum ferrimagnetic ground state \eqref{gs0} at zero magnetic field
\begin{eqnarray}
{\cal N} \!\!\!&=&\!\!\! \frac{\sqrt{(1\!-\!2\frac{D}{J})^2 \!+\! 8\Delta^2} - (1\!-\!2\frac{D}{J})}{4 \sqrt{(1\!-\!2\frac{D}{J})^2 \!+\! 8\Delta^2}}  \nonumber \\
         \!\!\!&\times&\!\!\! \left[\sqrt{\frac{5\sqrt{(1\!-\!2\frac{D}{J})^2 \!+\! 8\Delta^2} + 3 (1\!-\!2\frac{D}{J})}{\sqrt{(1\!-\!2\frac{D}{J})^2 \!+\! 8\Delta^2} 
				- (1\!-\!2 \frac{D}{J})}} - 1 \right]\!\!.
\label{eqneq0}
\end{eqnarray}
It appears worthwhile to examine in somewhat more detail a degree of the bipartite quantum entanglement of the mixed spin-(1/2,1) Heisenberg dimer in zero magnetic field depending on a relative strength of the exchange and uniaxial single-ion anisotropy. To this end, the negativity is plotted in Fig.~\ref{fig1} against the uniaxial single-ion anisotropy for three representative values of the exchange anisotropy, namely, the fully isotropic case ($\Delta=1.0$), the particular case with the easy-axis ($\Delta=0.5$) and easy-plane ($\Delta=2.0$) exchange anisotropy. It is evident from Fig.~\ref{fig1} that the negativity shows a relatively broad maximum at ${\cal N} = 1/3$, whose position depends on a specific choice of the exchange anisotropy. For instance, the mixed spin-(1/2,1) Heisenberg dimer with a perfectly isotropic exchange interaction $\Delta = 1$ exhibits the strongest bipartite quantum entanglement on assumption that the uniaxial single-ion anisotropy is also absent $D/J = 0$, i.e., it does not possess any form of the magnetic anisotropy. On the other hand, the easy-axis (easy-plane) exchange anisotropy $\Delta < 1$ ($\Delta >1$) shifts the local maximum of the negativity towards the uniaxial single-ion anisotropy with an easy-plane (easy-axis) character $D/J > 0$ ($D/J < 0$) competing with the exchange anisotropy. It is also worthy to note that a relative strength of the bipartite quantum entanglement becomes according to Eq. \eqref{eqneq0} independent of the exchange anisotropy for the particular value of the uniaxial single-ion anisotropy $D/J = 1/2$, for which the negativity acquires a half of the golden-ratio conjugate ${\cal N} = (\sqrt{5} - 1)/4 \approx 0.309$ regardless of the anisotropy parameter $\Delta$ (see a crossing point in Fig.~\ref{fig1}).

\begin{figure}[t!]
{\includegraphics[width=.49\textwidth,trim=3cm 9.5cm 3.5cm 9.5cm, clip]{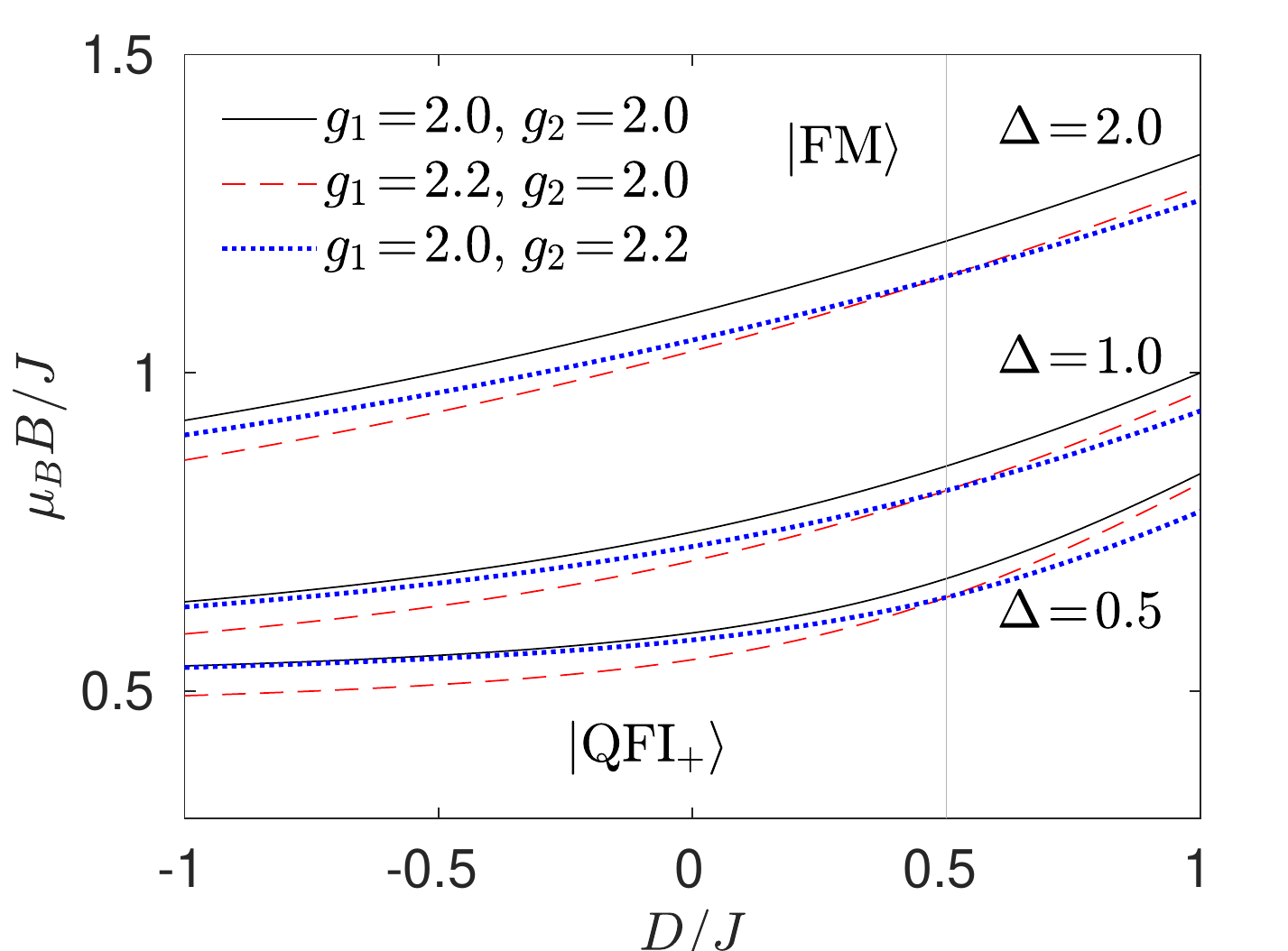}}
\vspace{-0.3cm}
\caption{The ground-state phase diagram in the $D/J$-$\mu_BB/J$ plane for three selected values of the exchange anisotropy $\Delta=0.5,1.0,2.0$, and three different sets of the Land\'e  $g$-factors indicated in the legend. A thin vertical line at $D/J=1/2$ determines a special value of the uniaxial single-ion anisotropy, at which the respective transition field becomes independent of the difference $|g_1-g_2|$.}
\label{fig2}
\end{figure}

Next, let us investigate in detail a strength of the bipartite quantum entanglement of the mixed spin-(1/2,1) Heisenberg dimer at zero temperature for three considered settings of Land\'e $g$-factors in presence of the external magnetic field. For this purpose, we have plotted first in Fig.~\ref{fig2} the ground-state phase diagrams of the mixed spin-(1/2,1) Heisenberg dimer in the $D/J$-$\mu_BB/J$ plane for three different values of the exchange anisotropy $\Delta=0.5$, $1.0$, and $2.0$. Of course, the sufficiently strong magnetic field gives rise to the classical ferromagnetic state $\left\vert \mbox{FM} \right\rangle=\left\vert1/2,1\right\rangle$, which is naturally without any quantum entanglement as evidenced by a zero value of the negativity ${\cal N}=0$. On the other hand, the two-fold degeneracy of the quantum ferrimagnetic ground state \eqref{gs0} is lifted by the external magnetic field due to the Zeeman splitting of energy levels, which stabilizes at sufficiently low but nonzero magnetic fields the unique quantum ferrimagnetic ground state  
\begin{eqnarray}
| {\rm QFI}_{+} \rangle = c_1^{+} | \!-\!1/2, 1 \rangle - c_1^{-} | 1/2, 0 \rangle. 
\label{gsf}
\end{eqnarray}
The unique quantum ferrimagnetic ground state \eqref{gsf} is characterized by a quantum superposition of the microstates $\left\vert-1/2,1\right\rangle$ and $\left\vert1/2,0\right\rangle$ unambiguously given by the following probability amplitudes
\begin{eqnarray}
c_1^{\pm} = \frac{1}{\sqrt{2}} \sqrt{1\!\pm\!\frac{1\!-\!2 \frac{D}{J}\!-\!2 \frac{\mu_B B}{J} (g_1\!-\!g_2)}{\sqrt{\left[1\!-\!2\frac{D}{J}\!-\!2 \frac{\mu_B B}{J}(g_1\!-\!g_2)\right]^2\!+\!8 \Delta^2}}
}\;.
\label{pagsf}
\end{eqnarray}
The respective zero-temperature asymptotic value of the negativity for the nondegenerate quantum ferrimagnetic ground state $| {\rm QFI}_{+} \rangle$ given by Eq. \eqref{gsf} can be acquired by making use of the density operator $\hat{\rho} = \left\vert {\rm QFI}_{+} \right\rangle \left\langle {\rm QFI}_{+}\right\vert$  
\allowdisplaybreaks
\begin{align}
{\cal N}&=\frac{\sqrt{2}\Delta}{\sqrt{\left[1-2\frac{D}{J}-2(g_1-g_2) \frac{\mu_B B}{J} \right]^2 + 8\Delta^2}}.
\label{eq18}
\end{align} 
At fixed values of the model parameters, the specific value of the negativity \eqref{eq18} pertinent to the nondegenerate quantum ferrimagnetic ground state $| {\rm QFI}_{+} \rangle$ is surprisingly much greater at nonzero magnetic fields than the zero-field value \eqref{eqneq0} inherent to the two-fold degenerate quantum ferrimagnetic phase $| {\rm QFI}_{\pm} \rangle$ forming the respective ground state in the zero-field limit. It could be thus concluded that the Zeeman splitting of energy levels due to the external magnetic field leads to an unexpected sudden rise of the bipartite quantum entanglement of the mixed spin-(1/2,1) Heisenberg dimer, which is in contrast with naive expectation that the magnetic field suppresses the quantum entanglement. Moreover, it will be shown hereafter that the sudden rise of the bipartite entanglement due to rising magnetic field at absolute zero temperature is also preserved at sufficiently small but nonzero temperatures, which makes this feature especially interesting with regard to possible experimental testing (see the part \ref{te}).  

\begin{figure*}[t!]
{\includegraphics[width=.30\textwidth,trim=3.2cm 9.8cm 5.5cm 9.5cm, clip]{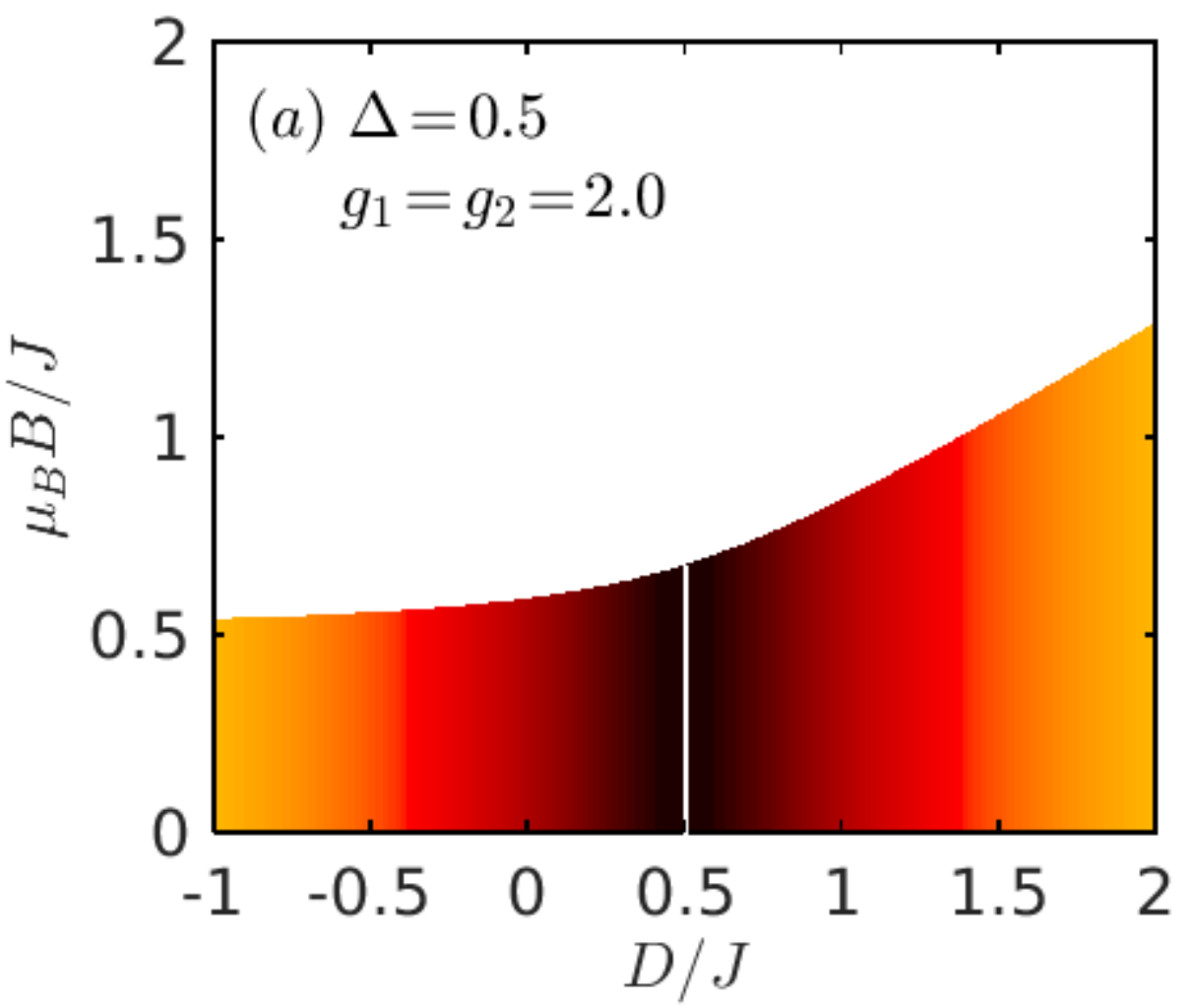}}
{\includegraphics[width=.30\textwidth,trim=3.2cm 9.8cm 5.5cm 9.5cm, clip]{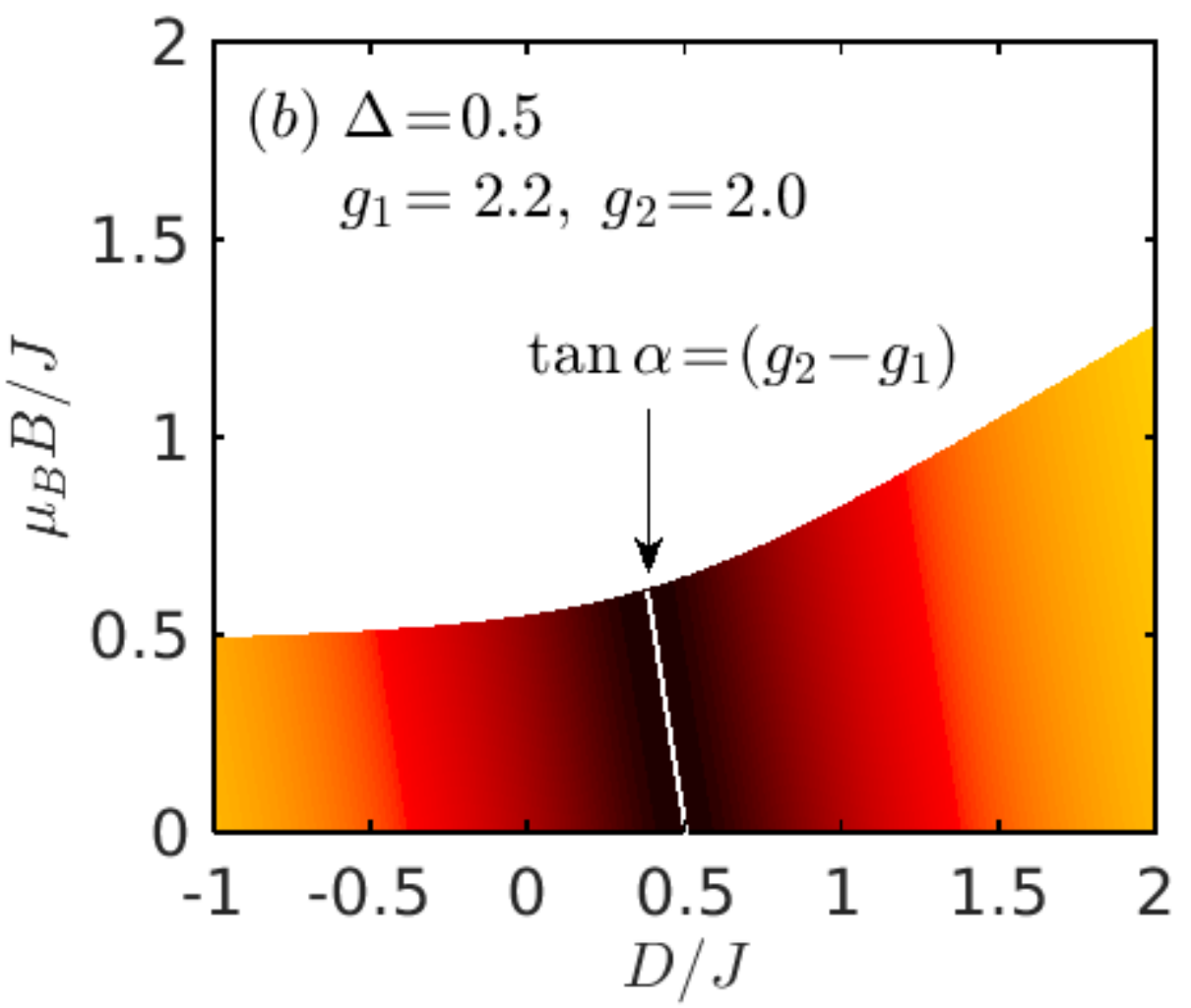}}
{\includegraphics[width=.366\textwidth,trim=3.2cm 9.8cm 2.8cm 9.5cm, clip]{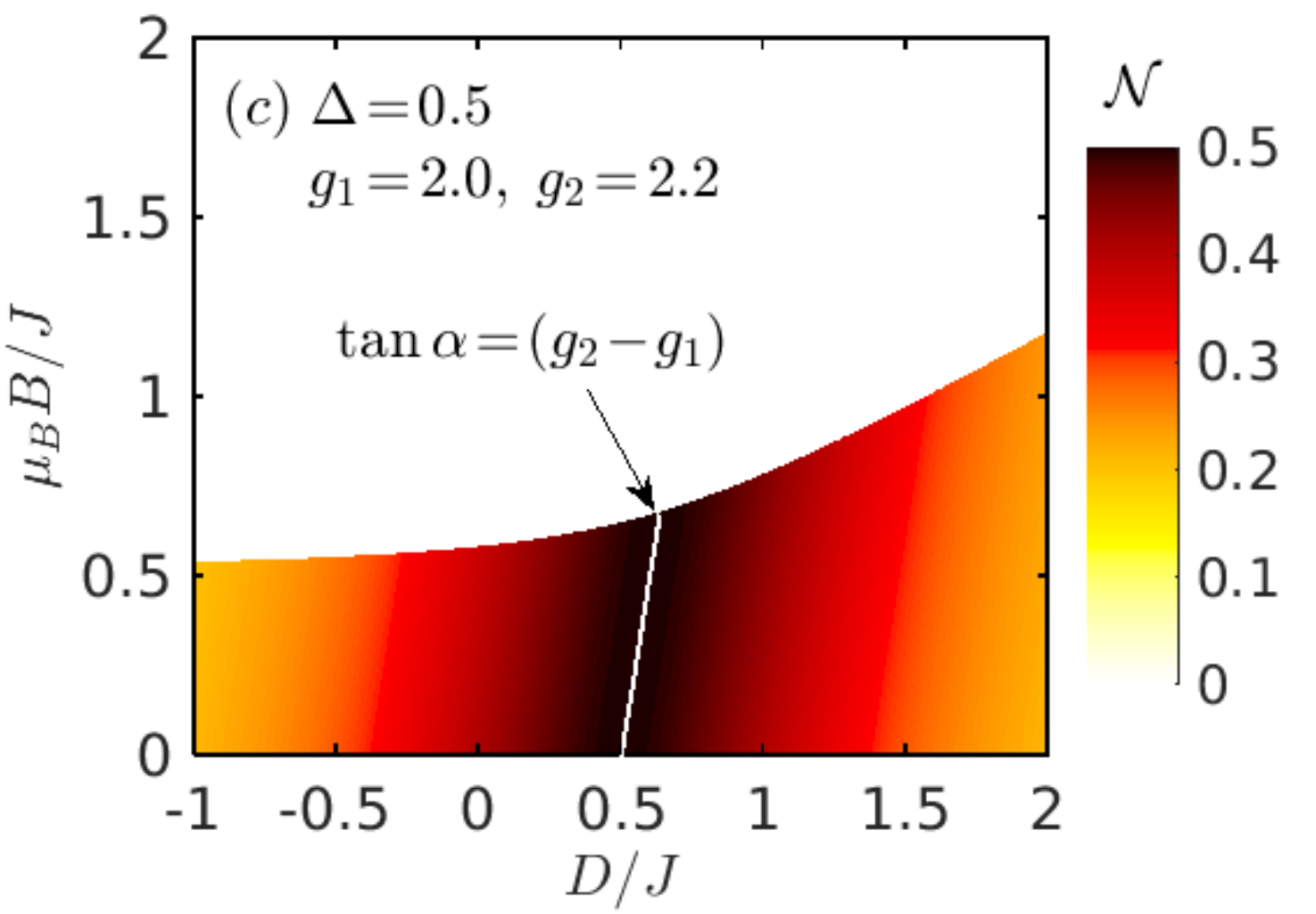}}\\
{\includegraphics[width=.30\textwidth,trim=3.2cm 9.8cm 5.5cm 9.5cm, clip]{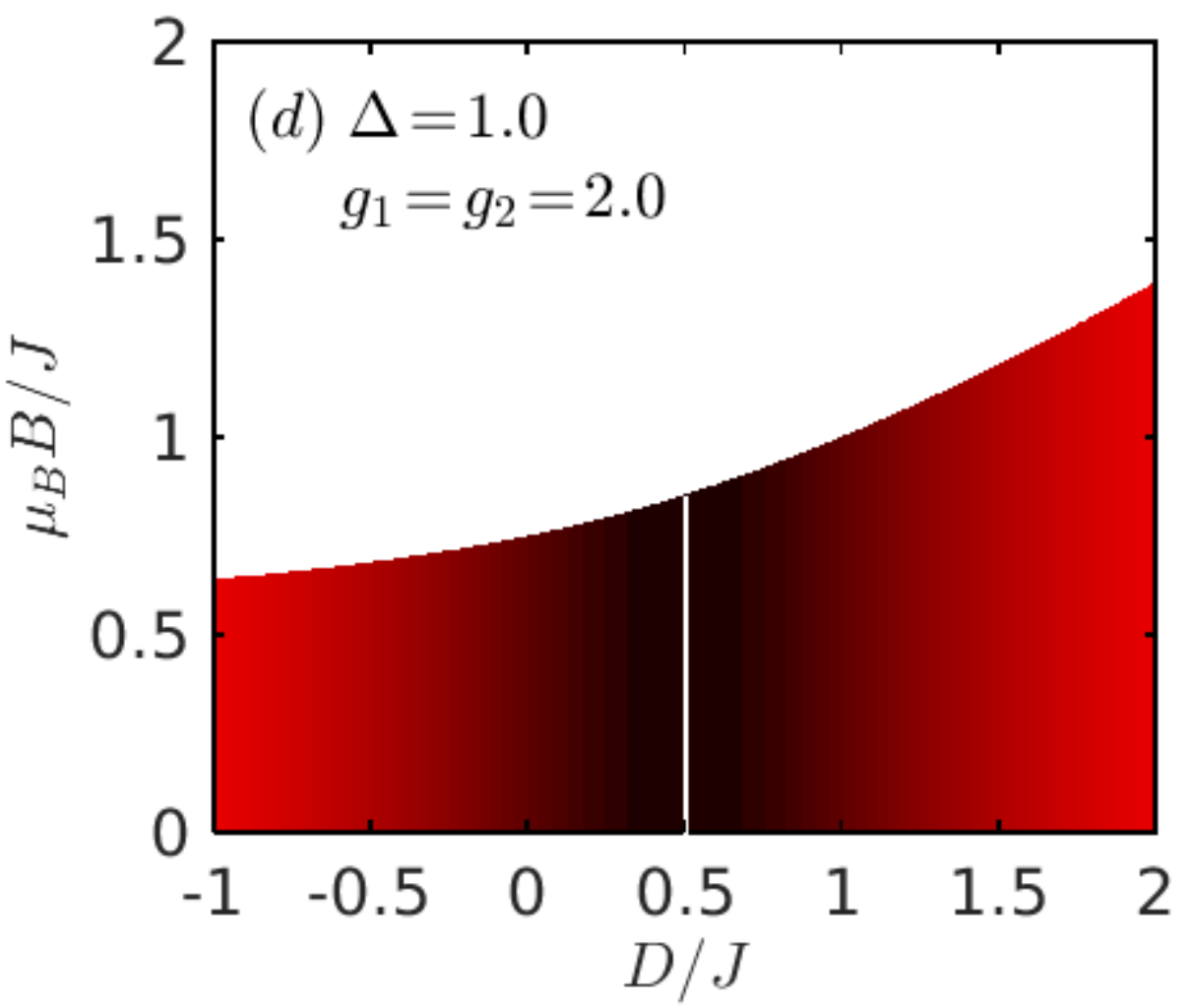}}
{\includegraphics[width=.30\textwidth,trim=3.2cm 9.8cm 5.5cm 9.5cm, clip]{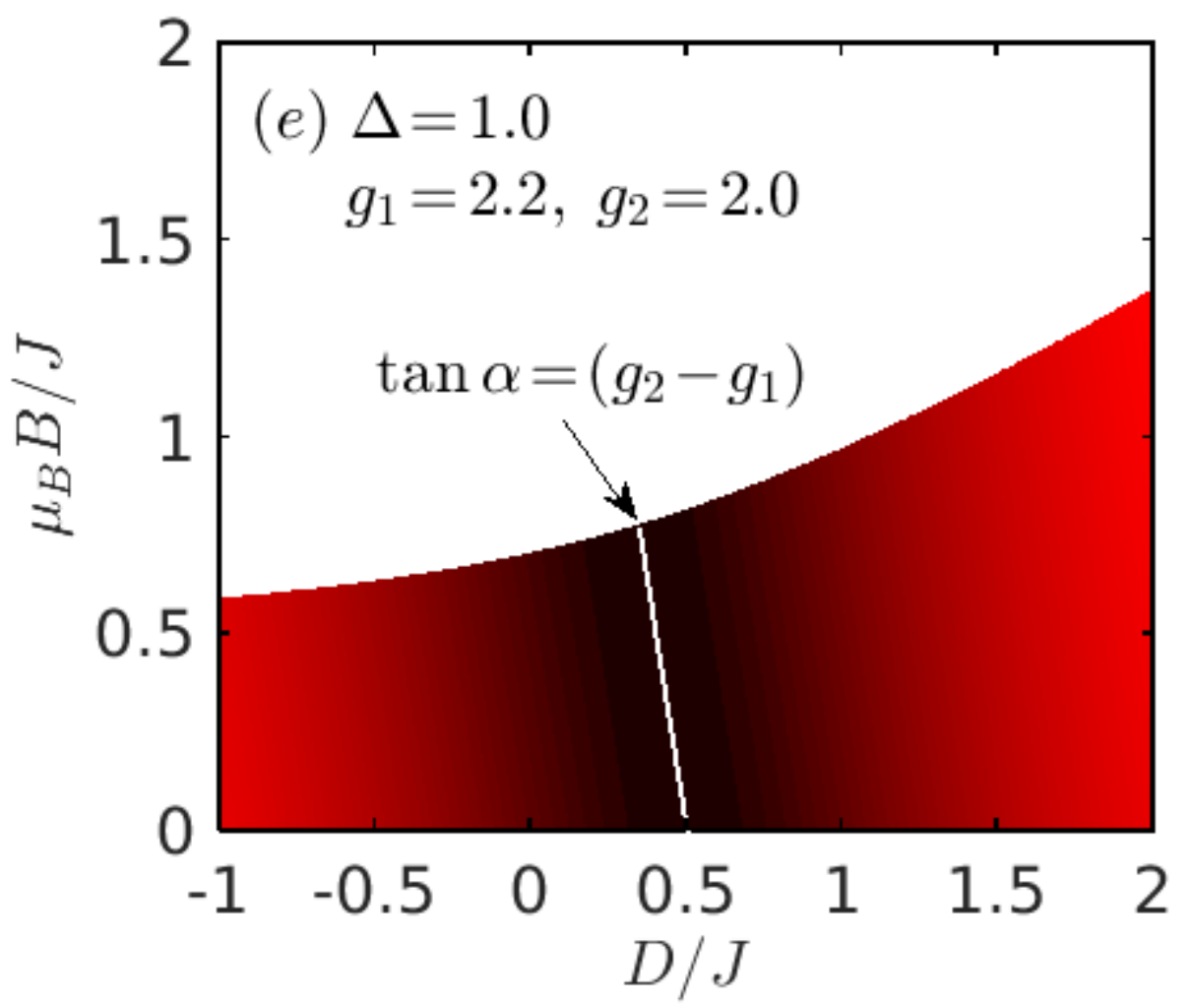}}
{\includegraphics[width=.366\textwidth,trim=3.2cm 9.8cm 2.8cm 9.5cm, clip]{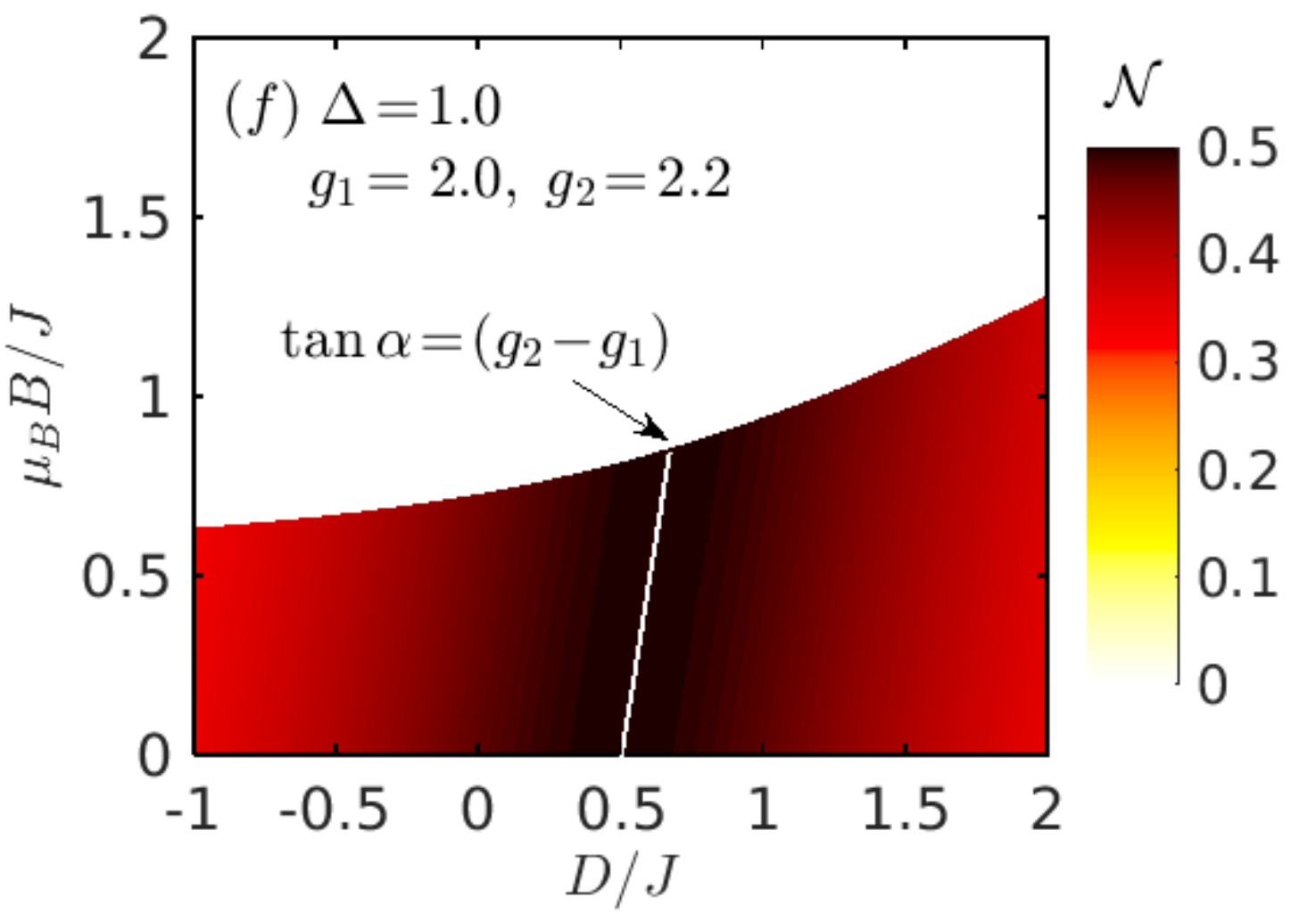}}
\caption{Zero-temperature density plots of the negativity ${\cal N}$ in the  $D/J$-$\mu_BB/J$ plane for three different sets of Land\'e $g$-factors specified in the panels and two selected values of the exchange anisotropy $\Delta = 0.5$ (upper panels) and $1.0$ (lower panels). A thin (white) line starting from $D/J = 1/2$ determines the contour line for the maximal value of the negativity ${\cal N} = 0.5$.}
\label{fig3}
\end{figure*}
 
The phase boundary between the classical ferromagnetic $\left\vert {\rm FM}\right\rangle$ and quantum ferrimagnetic $\left\vert {\rm QFI}_+\right\rangle$ ground states follows from the formula 
\begin{eqnarray}
\frac{\mu_BB}{J}=\frac{1}{4 g_1 g_2} \Biggl[g_1 \Biggr. \!\!\!&+&\!\!\! 2g_2 + 2g_1 \frac{D}{J} \label{eq19} \\
\!\!\!&+&\!\!\! \Biggl. \sqrt{\!\left(g_1 \!-\! 2g_2 \!+\! 2g_1 \frac{D}{J}\right)^{\!2} \!\!\!+\! 8 g_1 g_2 \Delta^{\!2}}\Biggr]\!,  \nonumber
\end{eqnarray} 
which depends on a mutual interplay of the exchange anisotropy $\Delta$, the uniaxial single-ion anisotropy $D/J$, as well as, the $g$-factors $g_1$ and $g_2$. In general, an increase of both anisotropy parameters $D/J$ and $\Delta$ stabilizes the quantum ferrimagnetic ground state $\left\vert {\rm QFI}_+\right\rangle$, while the rising magnetic field $\mu_B B/J$ contrarily stabilizes the classical ferromagnetic ground state $\left\vert {\rm FM}\right\rangle$. A shift of the gyromagnetic $g$-factors from their spin-only value also promotes existence of the classical ferromagnetic ground state $\left\vert {\rm FM}\right\rangle$ at the expense of the quantum ferrimagnetic ground state $\left\vert {\rm QFI}_+\right\rangle$, however, this impact is rather insignificant for reasonable values of gyromagnetic ratios $g_{1,2} \gtrsim 2$. In addition, it can be clearly seen from Fig.~\ref{fig2} that the ground-state phase boundaries for two particular cases with unequal g-factors cross each other at the special value of the uniaxial single-ion anisotropy $D/J=1/2$ when assuming the same value of the exchange anisotropy $\Delta$, because the transition field is in accordance with Eq. \eqref{eq19} independent of a difference of the gyromagnetic $g$-factors $|g_1-g_2|$.
Zero-temperature density plots of the negativity ${\cal N}$, which quantifies a degree of the bipartite quantum entanglement within the mixed spin-(1/2,1) Heisenberg dimer, are depicted in Fig.~\ref{fig3} in the $D/J$-$\mu_BB/J$ plane for three different sets of Land\'e $g$-factors and two representative values of the exchange anisotropy $\Delta=0.5$ and $1.0$. In agreement with the formula \eqref{eq18}, the negativity ${\cal N}$ becomes  within the quantum ferrimagnetic phase \eqref{gsf} fully independent of a relative strength of the magnetic field $\mu_BB/J$ on assumption that the gyromagnetic factors are set equal to each other $g_1=g_2$ (see left panels in Fig.~\ref{fig3}). Even under the specific constraint $g_1=g_2$, the anisotropic parameters $D/J$ and $\Delta$ still significantly influence a strength of the bipartite quantum entanglement, for instance, the negativity ${\cal N}$ is in general reinforced upon increasing of the parameter $\Delta$. As far as the influence of the uniaxial single-ion anisotropy is concerned, the maximal value of the negativity ${\cal N} = 0.5$ is notably reached for the particular case with $D/J=1/2$ (see vertical white lines in left panels of Fig.~\ref{fig3}), whereas the negativity gradually diminishes as one moves further apart from this specific case to the highly anisotropic cases $D/J\to\pm\infty$.

A situation for the more general case with different gyromagnetic $g$-factors $g_1 \neq g_2$ is much more involved. The contours with extremal values of the negativity in the middle and right panels of Fig.~\ref{fig3}, along which the negativity achieves the maximal value ${\cal N} = 0.5$, are apparently not vertical, but they deflect from a vertical direction by the specific angle $\alpha$ being proportional to a difference of Land\'e $g$-factors $\alpha= \arctan(g_2-g_1)$. It is noteworthy that the same trend is preserved also for contours, which do not correspond to the extremal value of the negativity. Owing to the inclination of the contours from the magnetic-field axis, the negativity may thus gradually increase or decrease upon variation of the magnetic field. Moreover, the increasing magnetic field may eventually initially enhance and successively reduce the negativity in the parameter region circumscribed by the specific values of the uniaxial single-ion anisotropy $D/J=1/2$ and $D/J=\left[2g_{2}-g_1-\sqrt{2}\Delta (g_{1}-g_2)\right]/2g_1$ before the field-driven transition between the quantum ferrimagnetic and classical ferromagnetic phase finally takes place.

\subsection{Thermal entanglement}
\label{te}

\begin{figure}[t!]
{\includegraphics[width=.9\columnwidth,trim=3cm 9.3cm 3.5cm 9.5cm, clip]{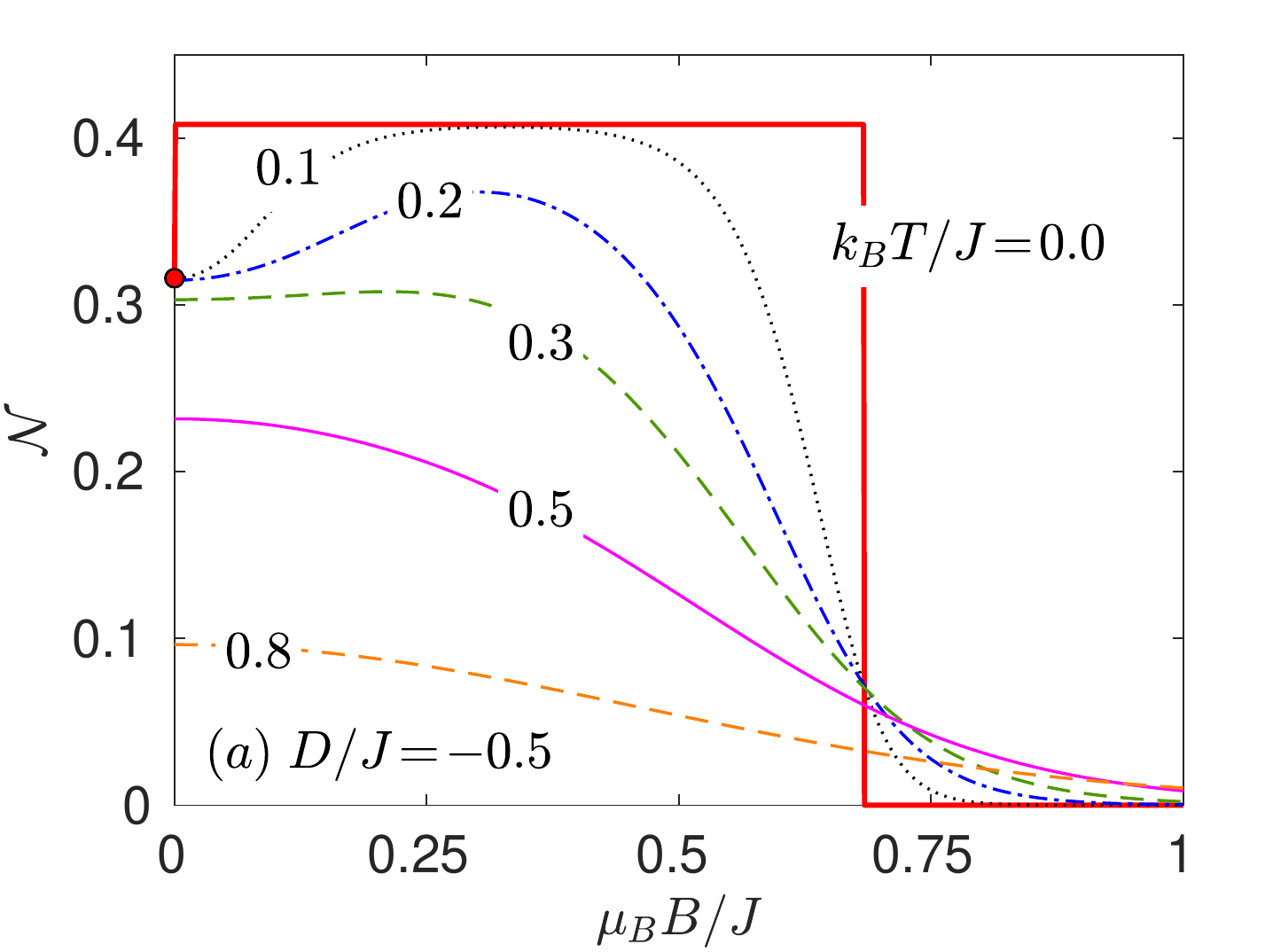}}
{\includegraphics[width=.9\columnwidth,trim=3cm 9.3cm 3.5cm 9.5cm, clip]{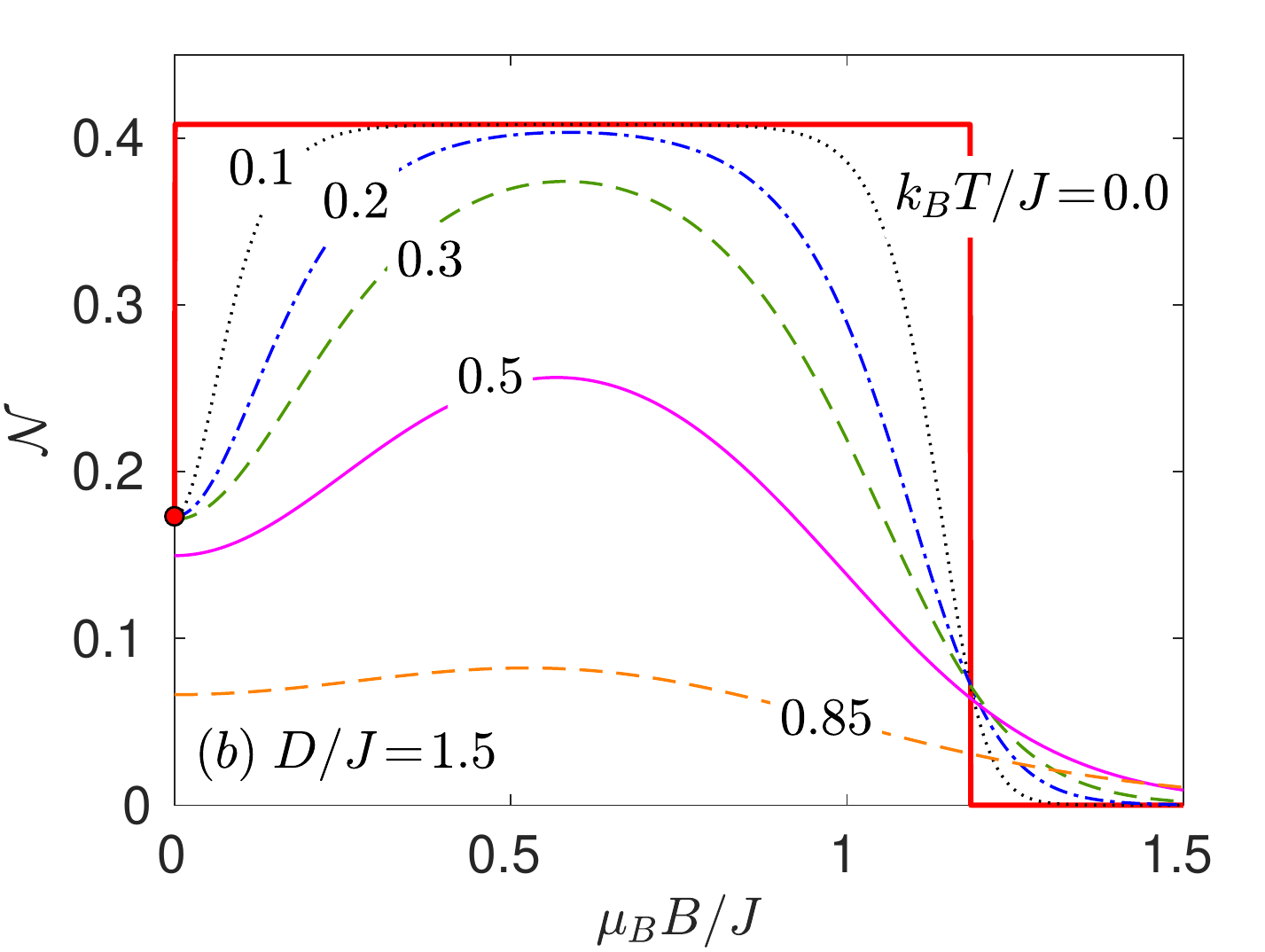}}
\caption{The negativity as a function of the magnetic field for $\Delta =1.0$, $g_1 = g_2 = 2.0$ by considering a few different values of temperature and two selected values of the uniaxial single-ion anisotropy: (a) $D/J = -0.5$; (b) $D/J = 1.5$. Red circled points determine a zero-field limit of the negativity for the absolute zero temperature.}
\label{fig4}
\end{figure}

Now, let us investigate in detail how the bipartite entanglement of the mixed spin-(1/2,1) Heisenberg dimer is resistant with respect to thermal fluctuations. The magnetic-field dependence of the negativity is shown in Fig.~\ref{fig4} for the specific case with $g_1=g_2$ and $\Delta=1.0$ at a few selected values of temperature and two different values of the uniaxial single-ion anisotropy, which are equally distant from the particular value $D/J=1/2$ bringing about the strongest quantum entanglement ${\cal N} = 0.5$. It is evident from Fig.~\ref{fig4} that the thermal entanglement is surprisingly enhanced upon increasing of the magnetic field as evidenced by a significant round maximum of the negativity observable for sufficiently low temperatures regardless of whether the uniaxial single-ion anisotropy is of easy-axis [Fig.~\ref{fig4}(a)] or easy-plane [Fig.~\ref{fig4}(b)] type. The unconventional enhancement of the thermal entanglement due to the magnetic field can be repeatedly related to Zeeman's splitting of two energy levels, which form the two-fold degenerate quantum ferrimagnetic ground state \eqref{gs0} in zero-field limit. As a matter of fact, the negativity converges at sufficiently low temperatures to the specific value \eqref{eq18}, which coincides with a degree of the quantum entanglement of the nondegenerate quantum ferrimagnetic ground state \eqref{gsf}.

\begin{figure*}[t!]
{\includegraphics[width=.2317\textwidth,trim=3.2cm 9.8cm 5.5cm 9.5cm, clip]{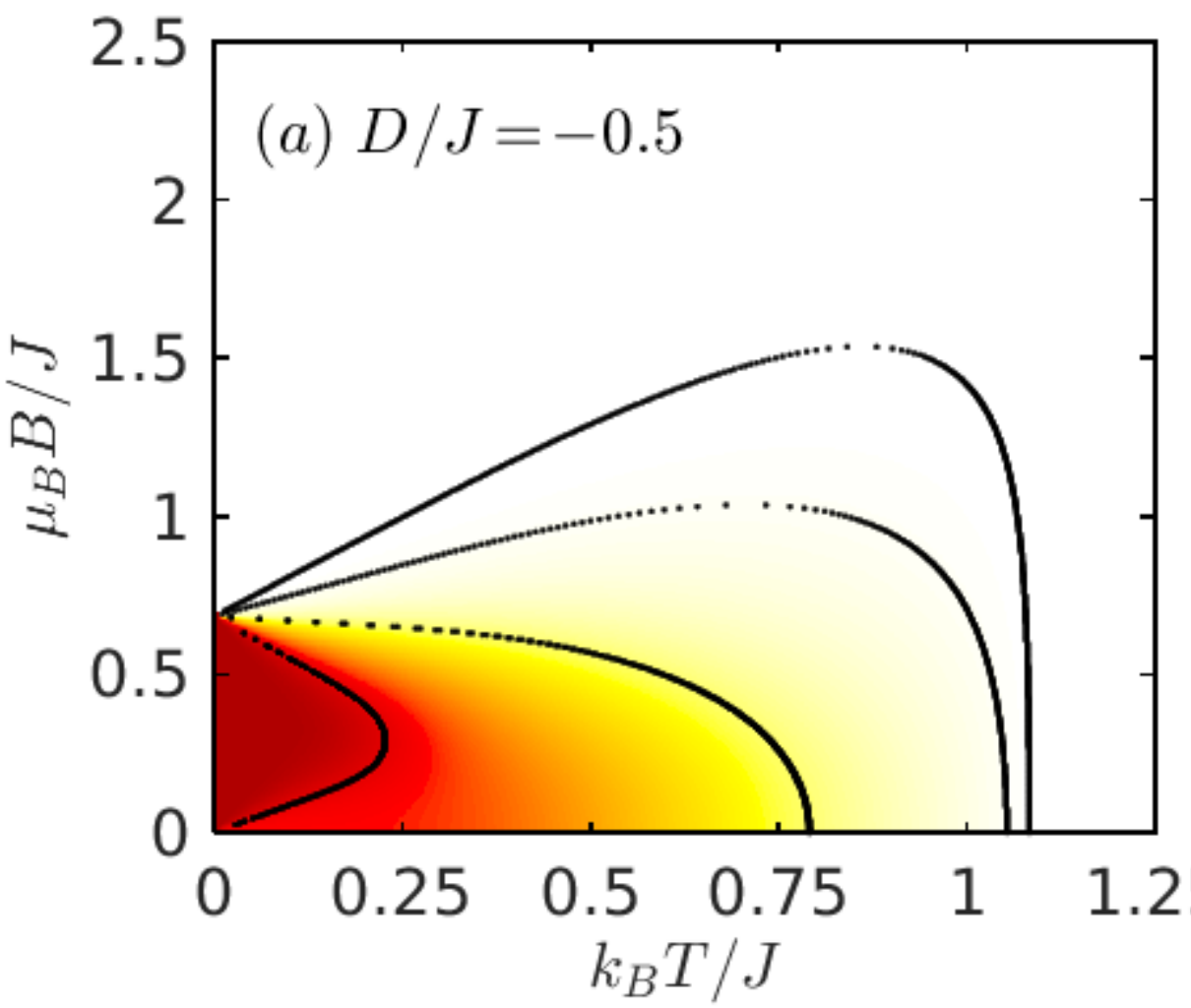}}
{\includegraphics[width=.2317\textwidth,trim=3.2cm 9.8cm 5.5cm 9.5cm, clip]{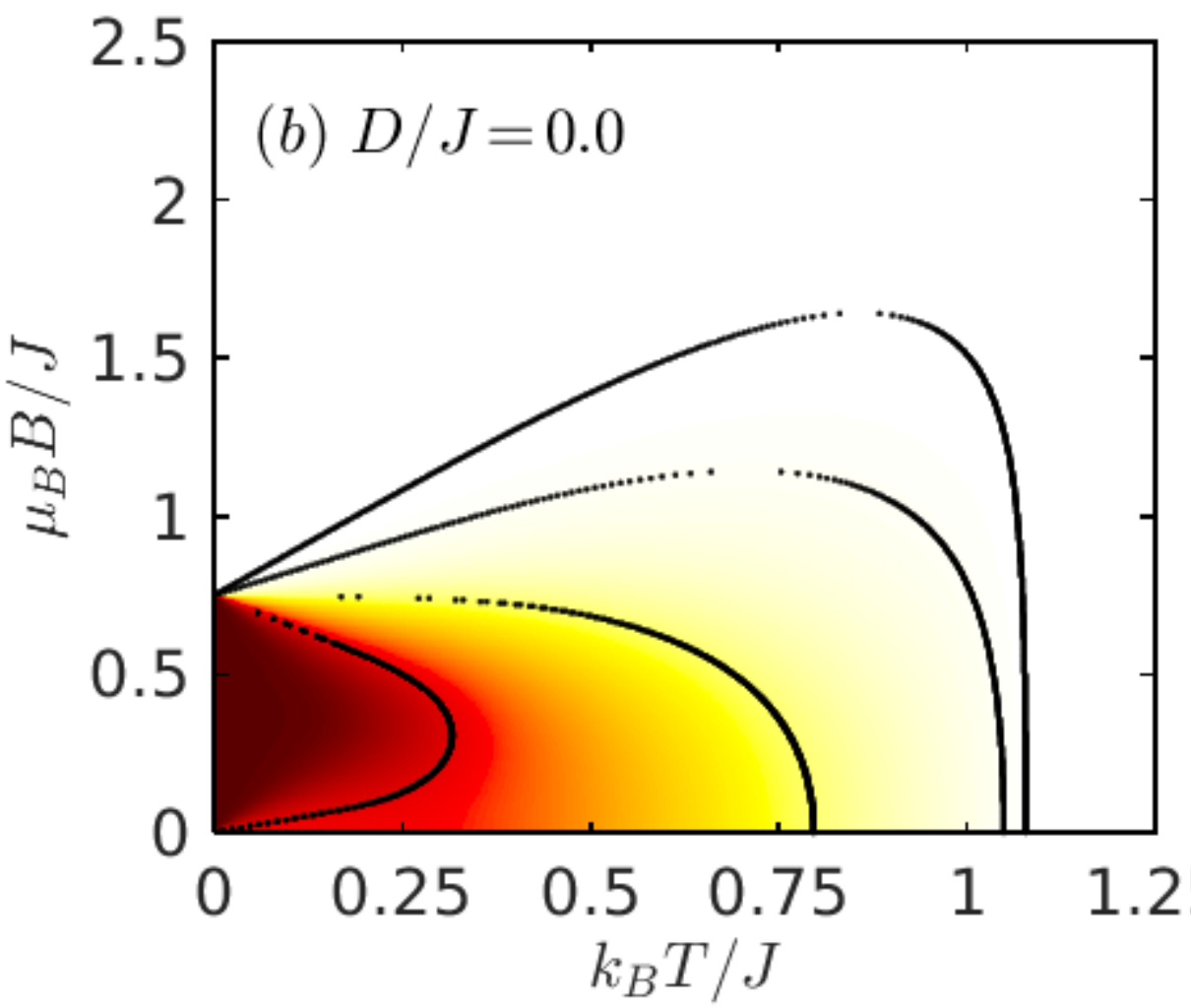}}
{\includegraphics[width=.2317\textwidth,trim=3.2cm 9.8cm 5.5cm 9.5cm, clip]{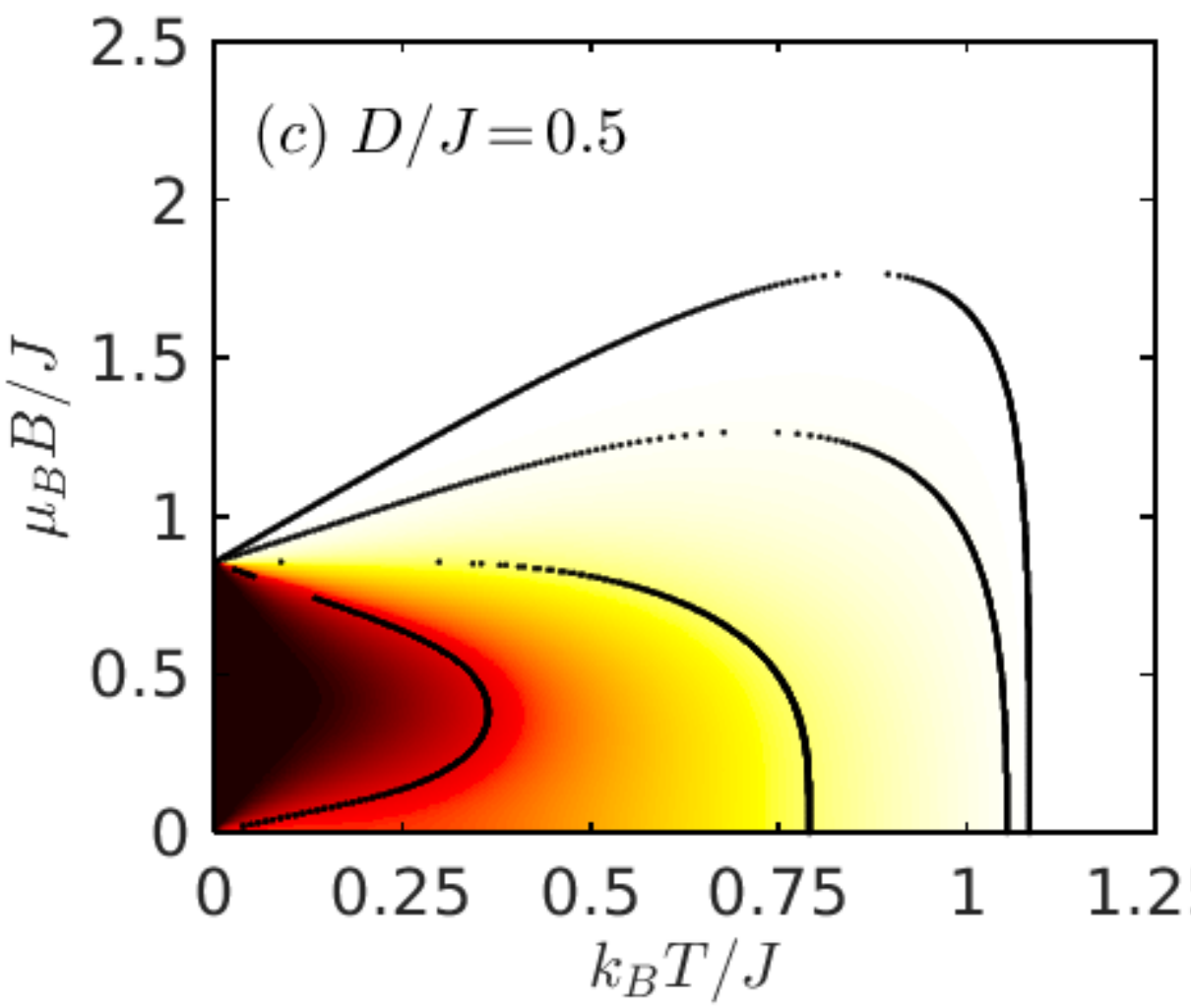}}
{\includegraphics[width=.288\textwidth,trim=3.2cm 9.8cm 2.8cm 9.5cm, clip]{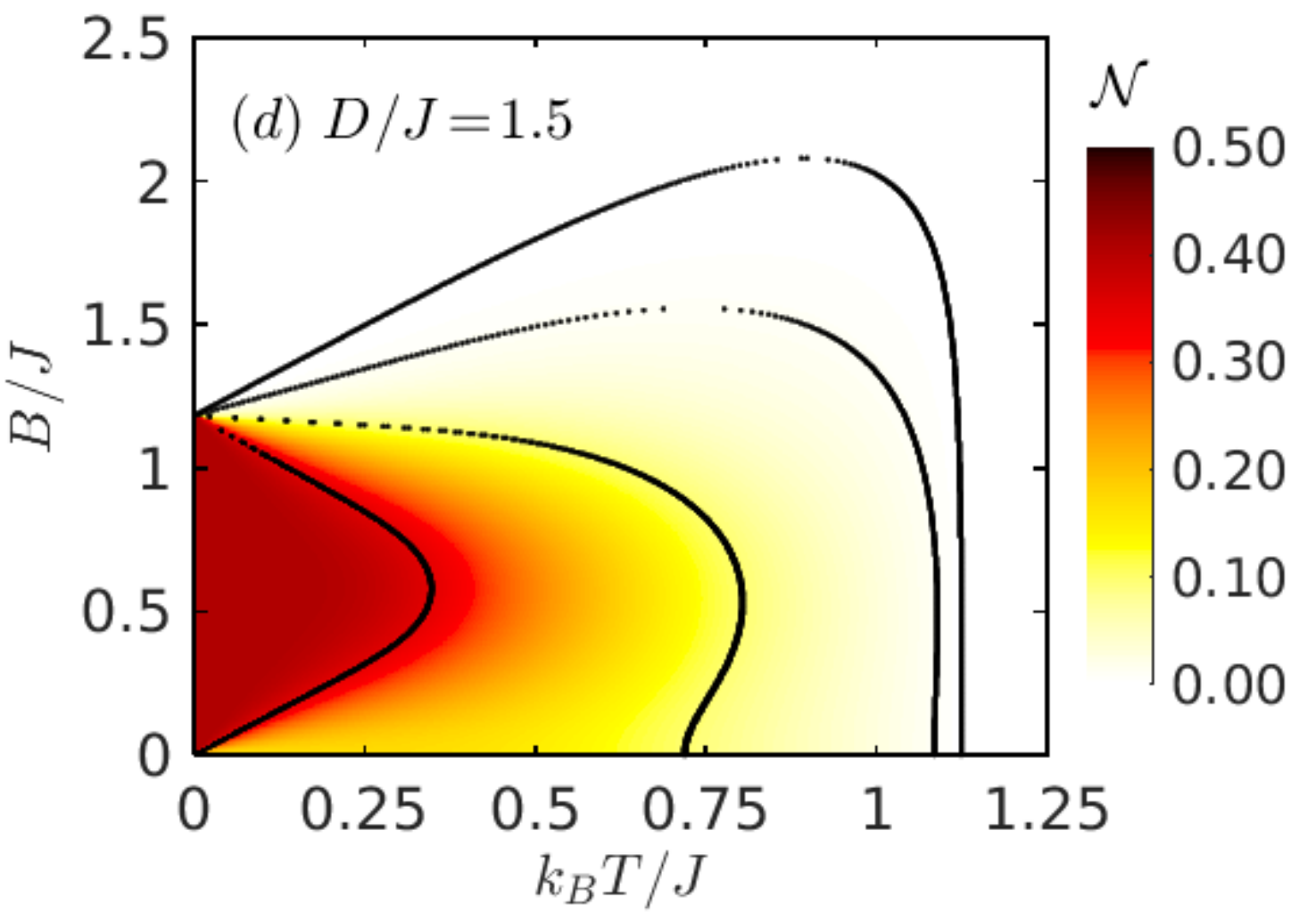}}
\caption{Density plots of the negativity ${\cal N}$ in the $k_BT/J$-$\mu_BB/J$ plane for $\Delta=1.0$, $g_1=g_2=2.0$ and four different values of the uniaxial single-ion anisotropy $D/J = -0.5$, $0.0$, $0.5$ and $1.5$. The black contour lines correspond to the specific values of ${\cal N}= 0.35$, 0.1, 0.01 and 0.001 (from left to right).}
\label{fig5}
\end{figure*}

To examine an influence of the uniaxial single-ion anisotropy on the thermal entanglement, a few density plots of the negativity ${\cal N}$ are displayed in Fig.~\ref{fig5} in the temperature-field plane by assuming the equal g-factors $g_1=g_2=2.0$, the fixed value of the exchange anisotropy $\Delta=1.0$ and four different values of the uniaxial single-ion anisotropy $D/J$.
It follows from the displayed density plots that a strong enough thermal entanglement can be detected only if temperature and magnetic field are simultaneously smaller than the exchange constant, i.e. $k_BT/J \lesssim 1$ and $\mu_BB/J \lesssim 1$. Moreover, the strongest thermal entanglement can be detected when the uniaxial single-ion anisotropy is sufficiently close to the specific value $D/J = 1/2$, which gives rise to the highest possible value of the negativity ${\cal N} = 0.5$ in the zero-temperature limit. It is also worth noticing that the thermal entanglement exhibits an intriguing reentrant behaviour when the external magnetic field is selected slightly above the saturation value. Under this condition, the negativity is initially zero at low enough temperatures, then it starts to develop above a lower threshold temperature until it reaches a local maximum and finally, the negativity gradually diminishes upon further increase of temperature until it completely disappears above an upper threshold temperature. A black contour line shown in Fig.~\ref{fig5} for the smallest value of the negativity indeed corroborates a temperature-driven reentrance of the thermal entanglement. In contrast to general expectations this result means that the relatively small thermal entanglement can be contraintuitively generated above the classical ferromagnetic ground state upon increasing of temperature.

\begin{figure*}[t!]
{\includegraphics[width=.45\textwidth,trim=3.cm 9.3cm 3.5cm 9.5cm, clip]{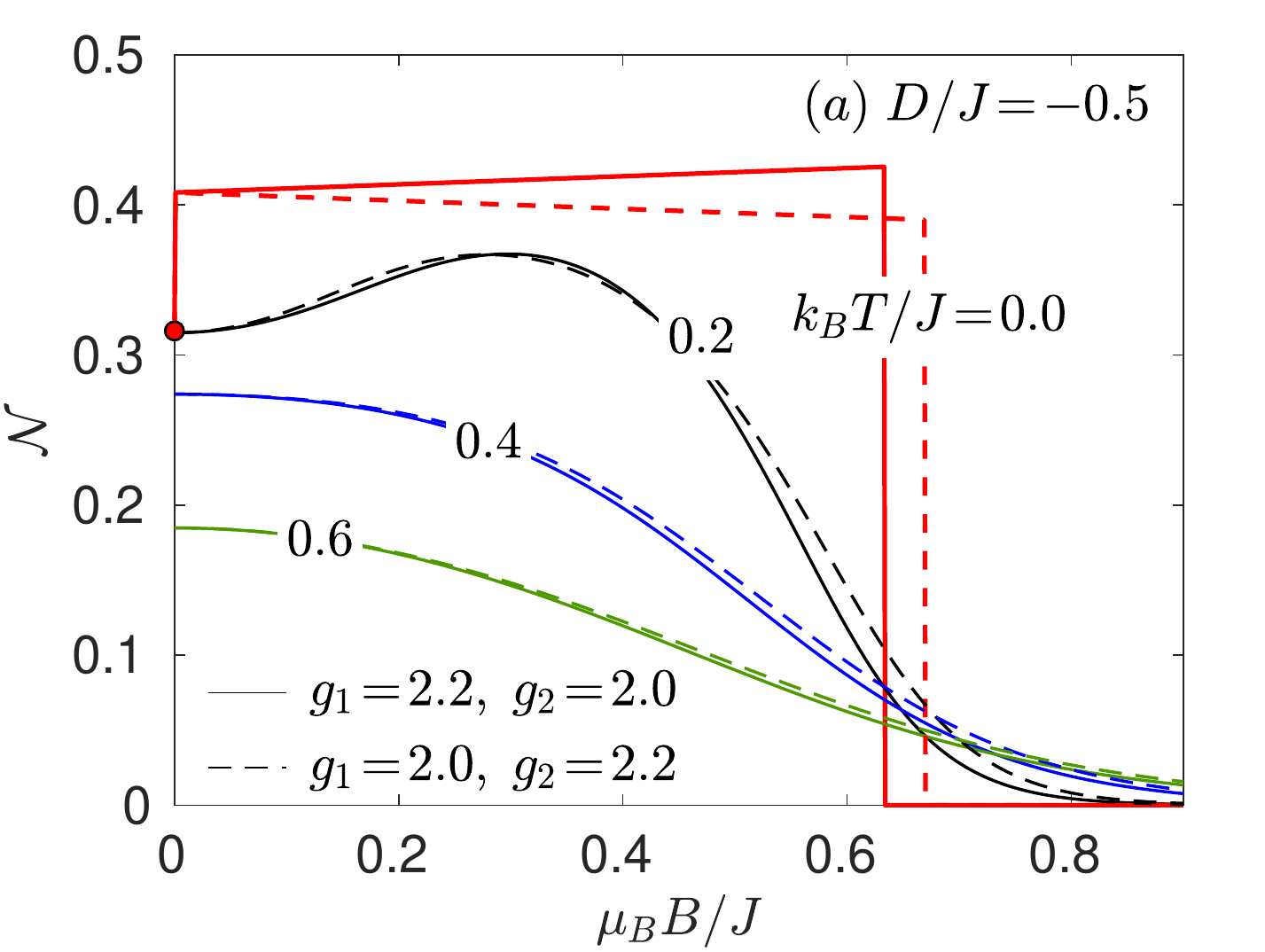}}
{\includegraphics[width=.45\textwidth,trim=3.cm 9.3cm 3.5cm 9.5cm, clip]{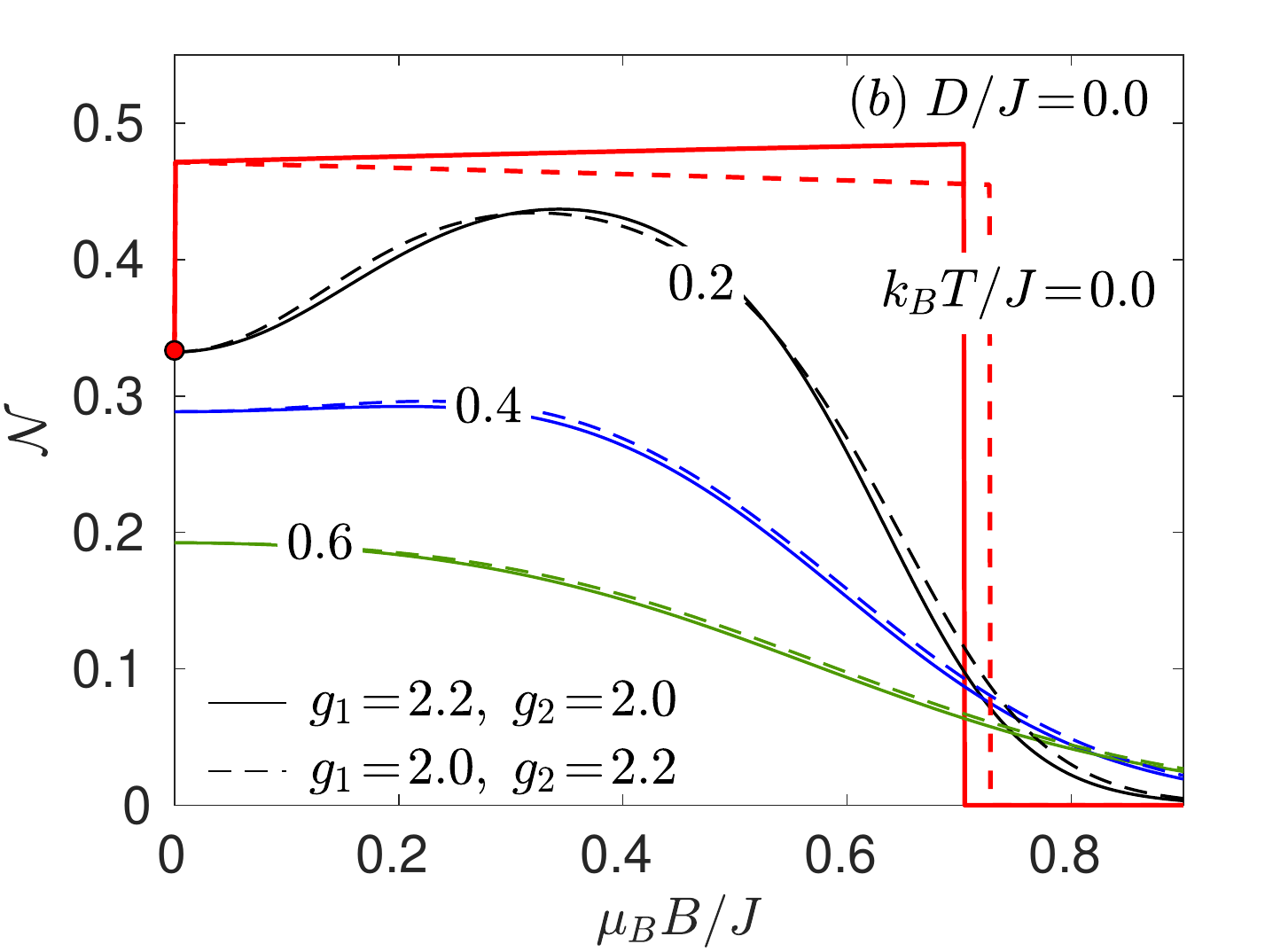}}
{\includegraphics[width=.45\textwidth,trim=3.cm 9.3cm 3.5cm 9.5cm, clip]{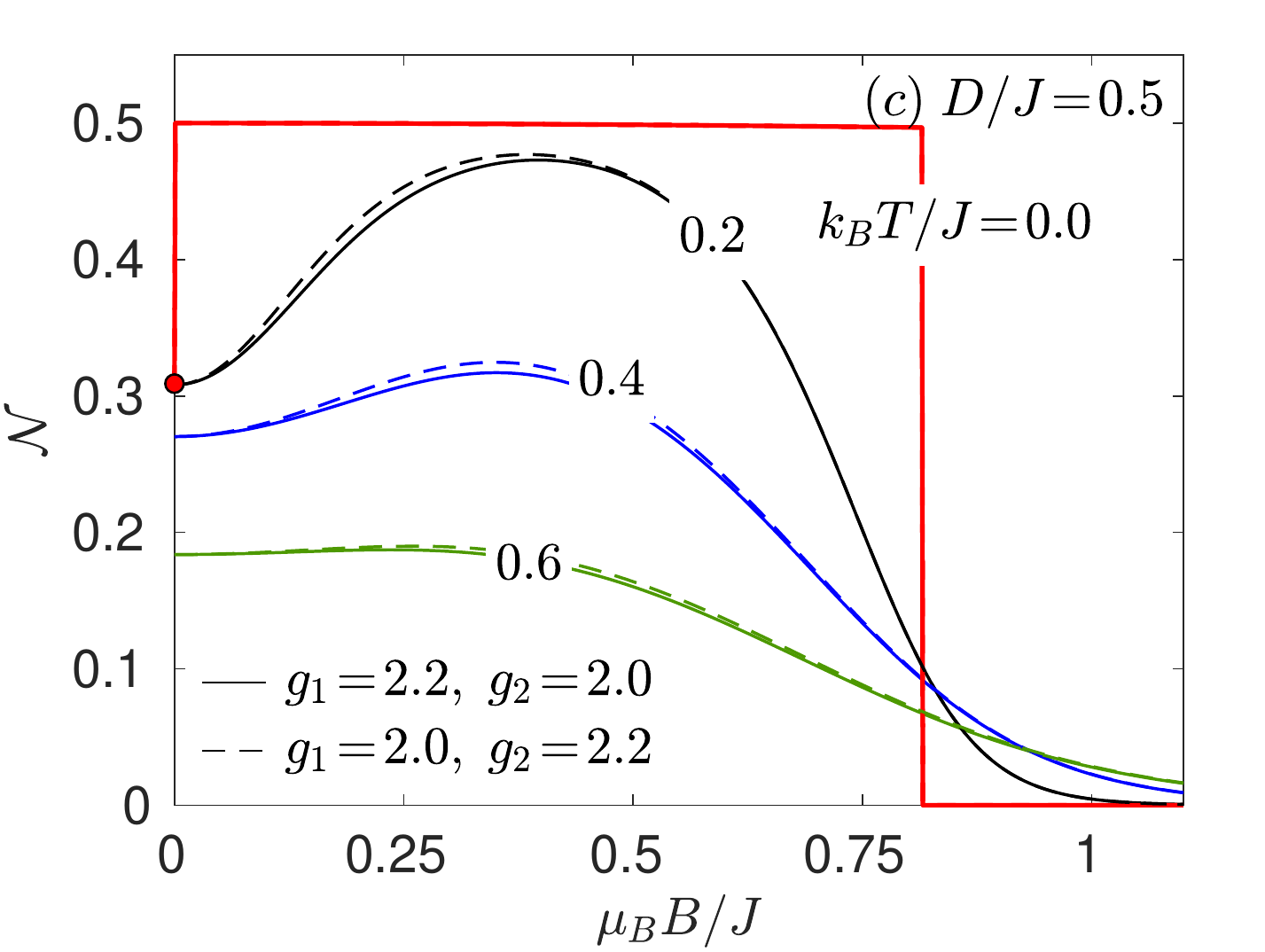}}
{\includegraphics[width=.45\textwidth,trim=3.cm 9.3cm 3.5cm 9.5cm, clip]{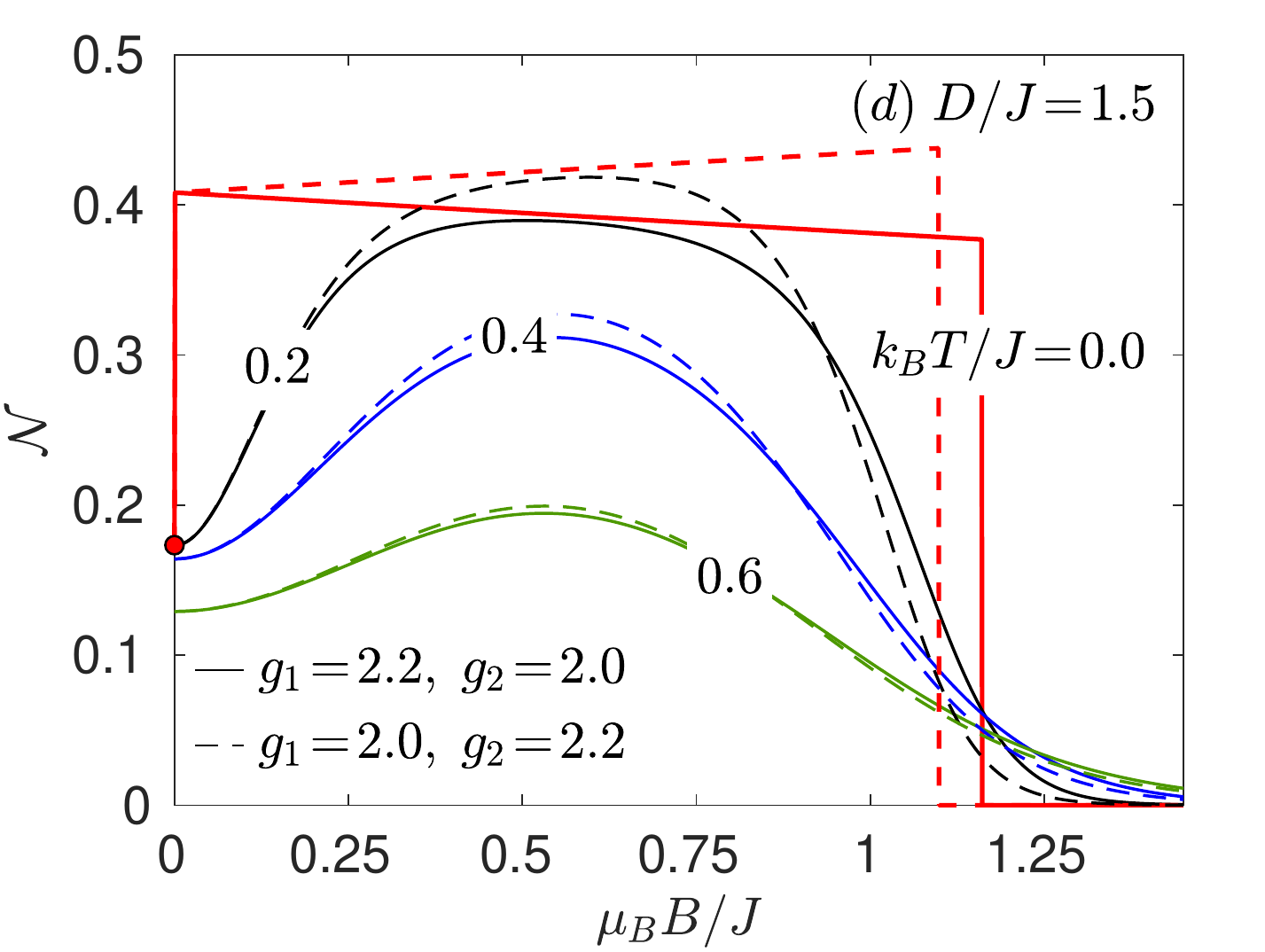}}
\caption{The negativity as a function of the magnetic field for the fixed value of the exchange anisotropy $\Delta=1.0$, a few different values of temperature, four selected vales of the uniaxial single-ion anisotropy $D/J = -0.5, 0.0, 0.5, 1.5$ and two different combination of the gyromagnetic factors with equal difference $|g_1-g_2|=0.2$.}
\label{fig6}
\end{figure*}

Next, our particular attention will be focused on how difference between the Land\'e g-factors may influence the thermal entanglement. To this end, the negativity ${\cal N}$ is plotted in Fig.~\ref{fig6} against the magnetic field for both types of differences of Land\'e $g$-factors $g_1>g_2$ and $g_1<g_2$, respectively, the fixed value of the exchange anisotropy $\Delta=1.0$, a few selected values of temperature $k_BT/J$ and four different values of the uniaxial single-ion anisotropy $D/J$.
Generally, the quantitative differences between the negativity for both considered settings of the gyromagnetic $g$-factors are very subtle mainly because of its small relative difference. It should be nevertheless pointed out that the negativity tends to the same asymptotic values in the zero-field limit as well as at high magnetic fields, while the most pronounced differences can be thus detected at moderate magnetic fields. It also follows from Fig.~\ref{fig6} that the zero-temperature asymptotic limit of the negativity shows in the low-field regime a quasi-linear increase (decrease) for $g_1>g_2$ on assumption that $D/J<1/2$ ($D/J > 1/2$), while the opposite trend applies for the other particular case with $g_1<g_2$. Most importantly, the negativity for $g_1<g_2$ mostly exceeds that one for $g_1>g_2$ even though the reverse statement may hold in a zero- and low-temperature limit.  

\begin{figure*}[t!]
{\includegraphics[width=.2317\textwidth,trim=3.2cm 9.8cm 5.5cm 9.5cm, clip]{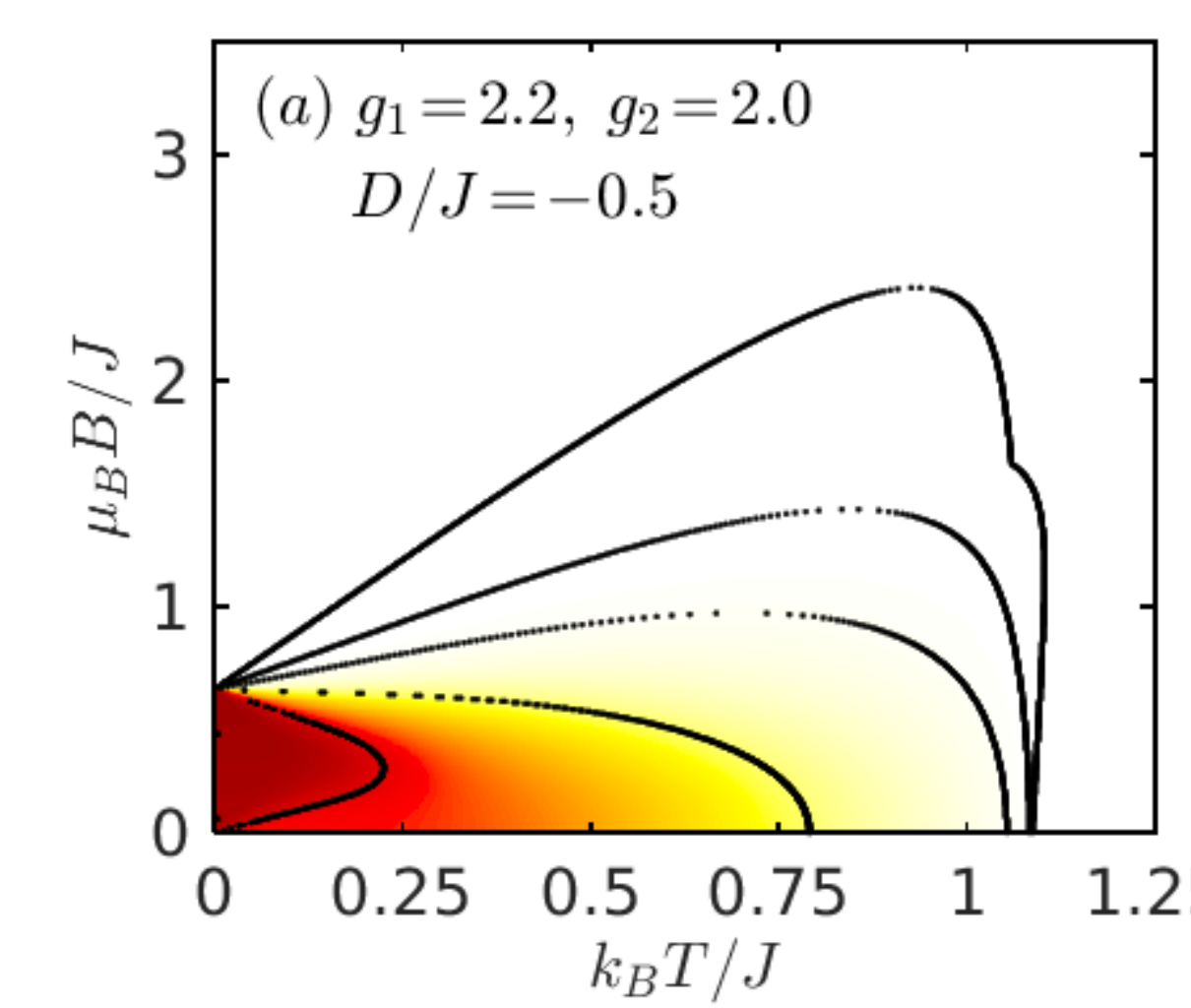}}
{\includegraphics[width=.2317\textwidth,trim=3.2cm 9.8cm 5.5cm 9.5cm, clip]{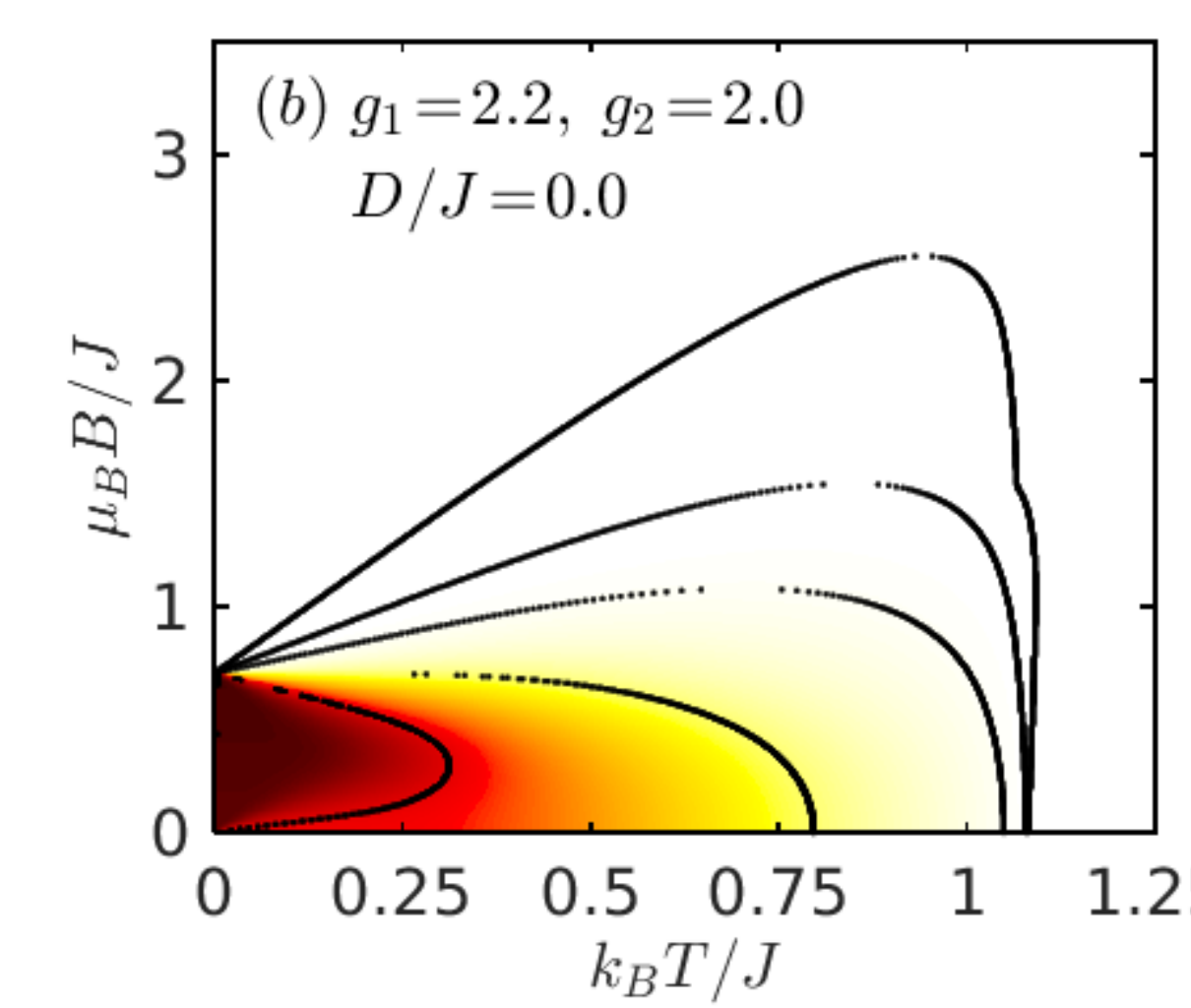}}
{\includegraphics[width=.2317\textwidth,trim=3.2cm 9.8cm 5.5cm 9.5cm, clip]{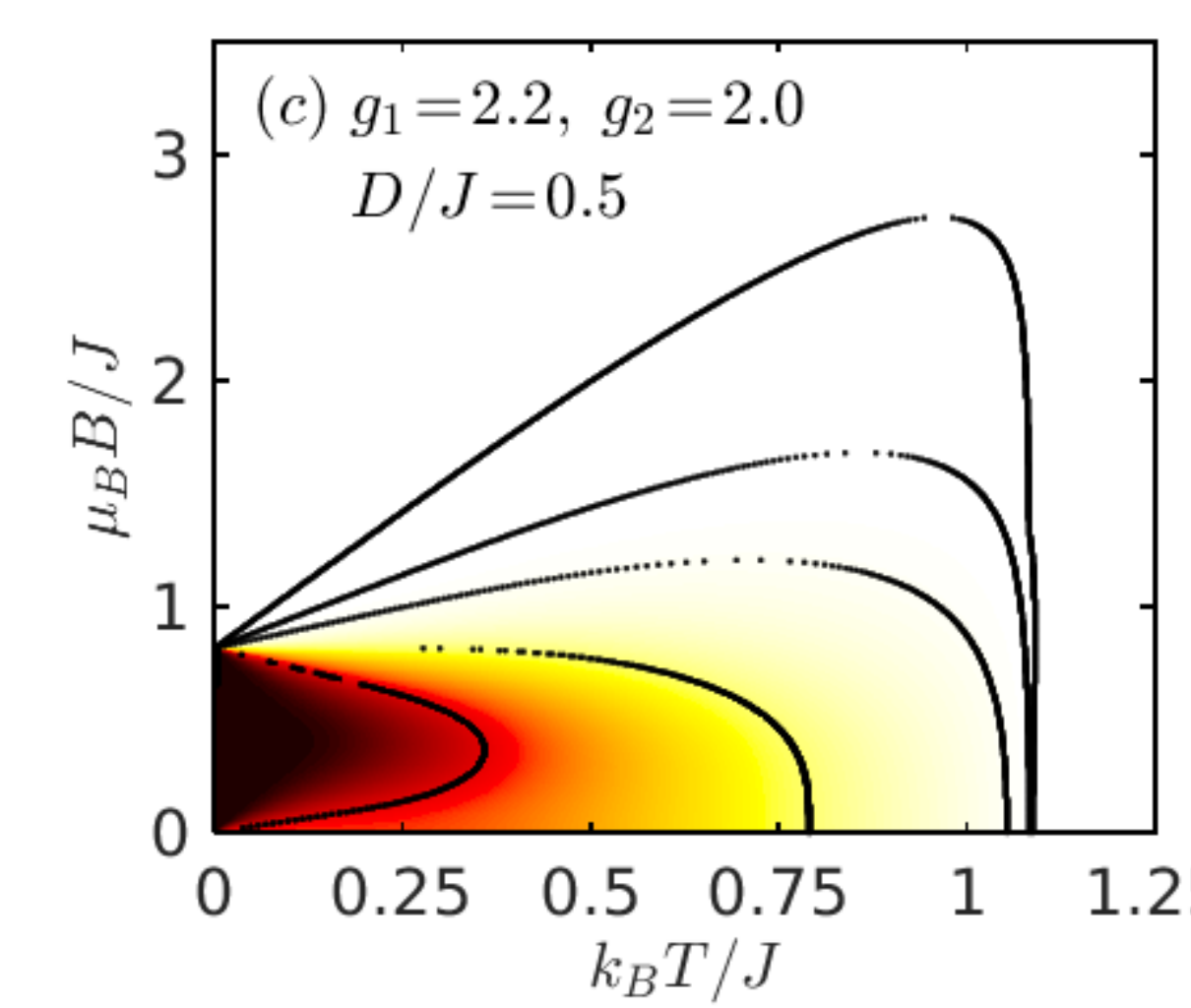}}
{\includegraphics[width=.288\textwidth,trim=3.2cm 9.8cm 2.8cm 9.5cm, clip]{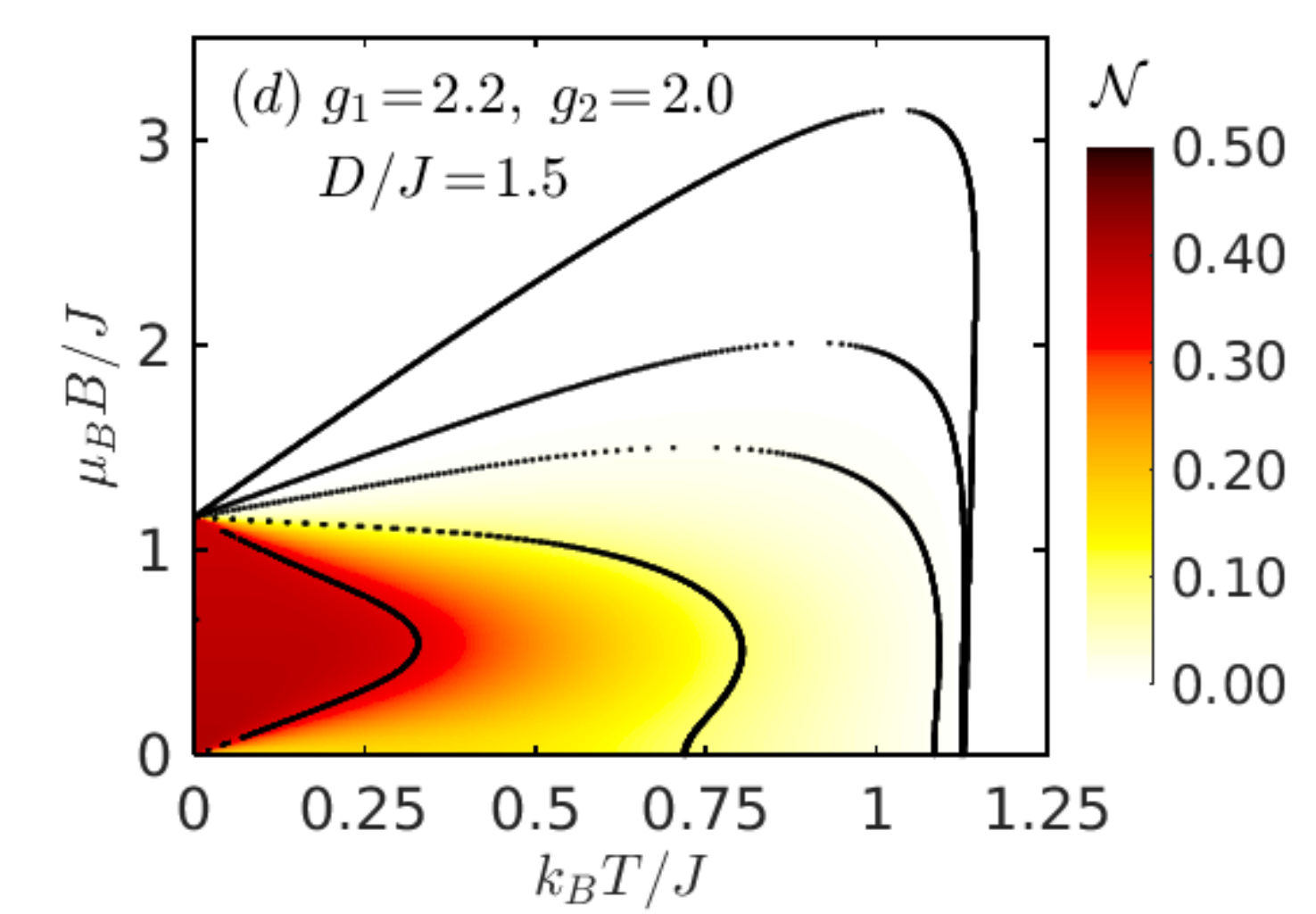}}\\
{\includegraphics[width=.2317\textwidth,trim=3.2cm 9.8cm 5.5cm 9.5cm, clip]{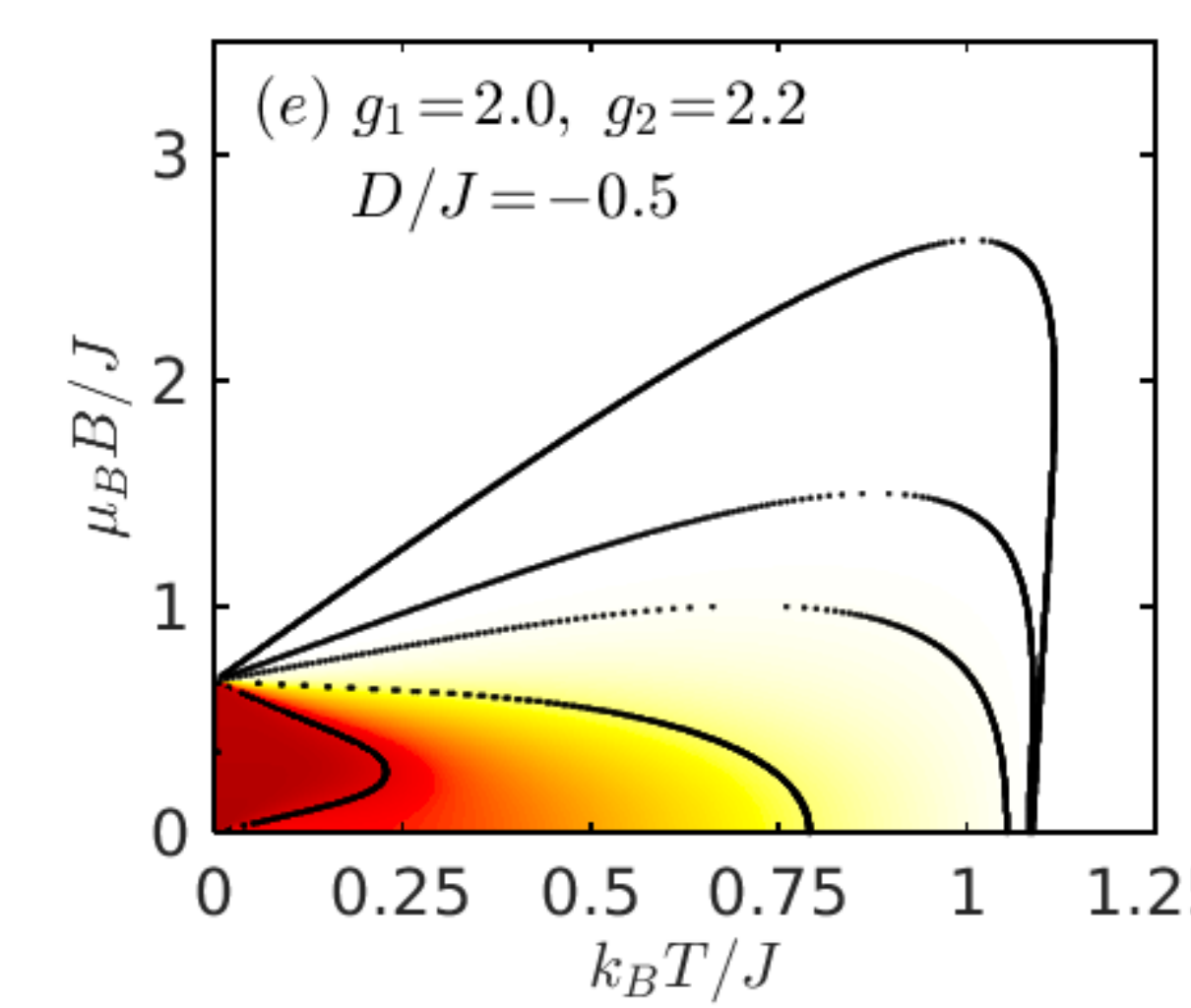}}
{\includegraphics[width=.2317\textwidth,trim=3.2cm 9.8cm 5.5cm 9.5cm, clip]{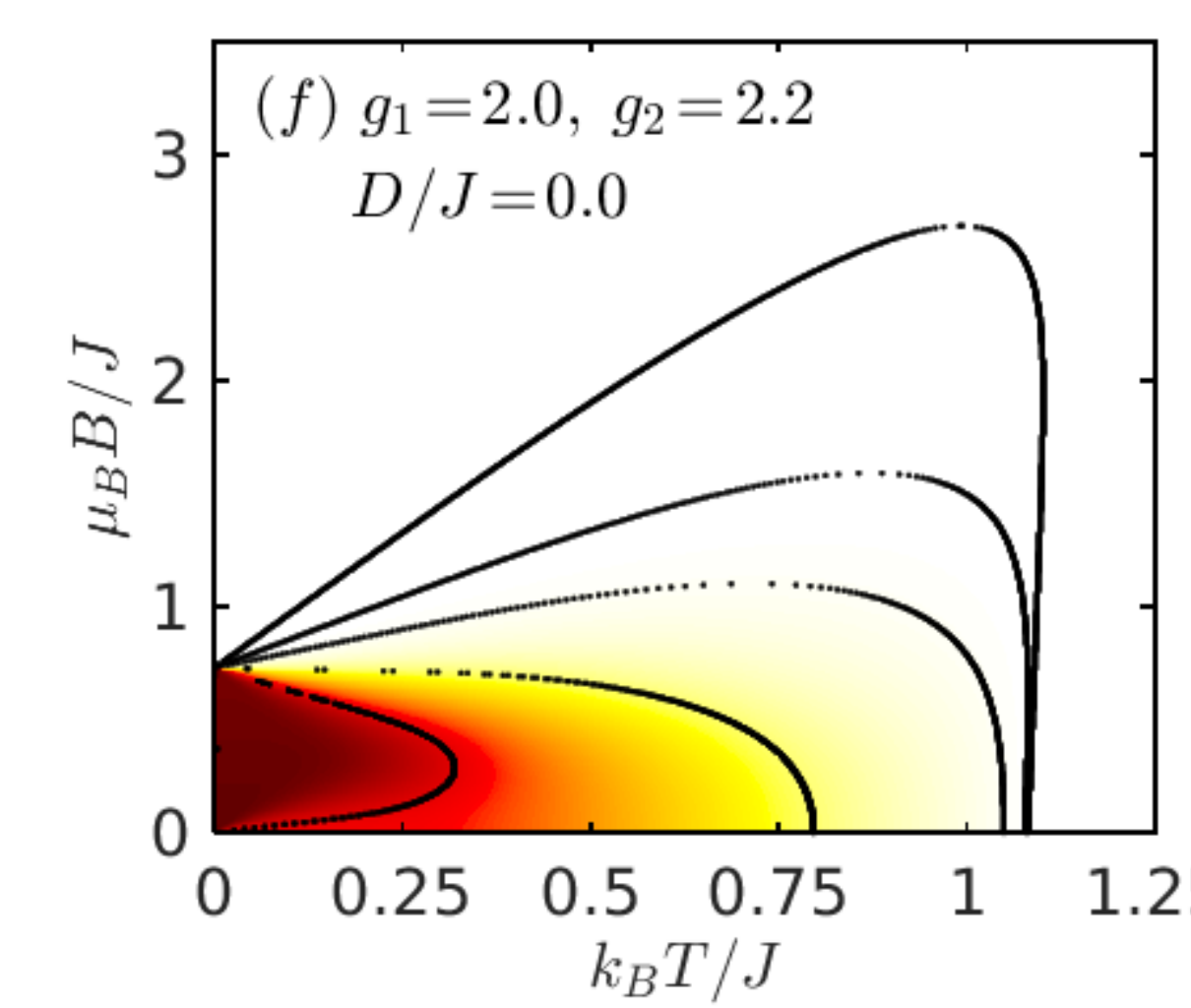}}
{\includegraphics[width=.2317\textwidth,trim=3.2cm 9.8cm 5.5cm 9.5cm, clip]{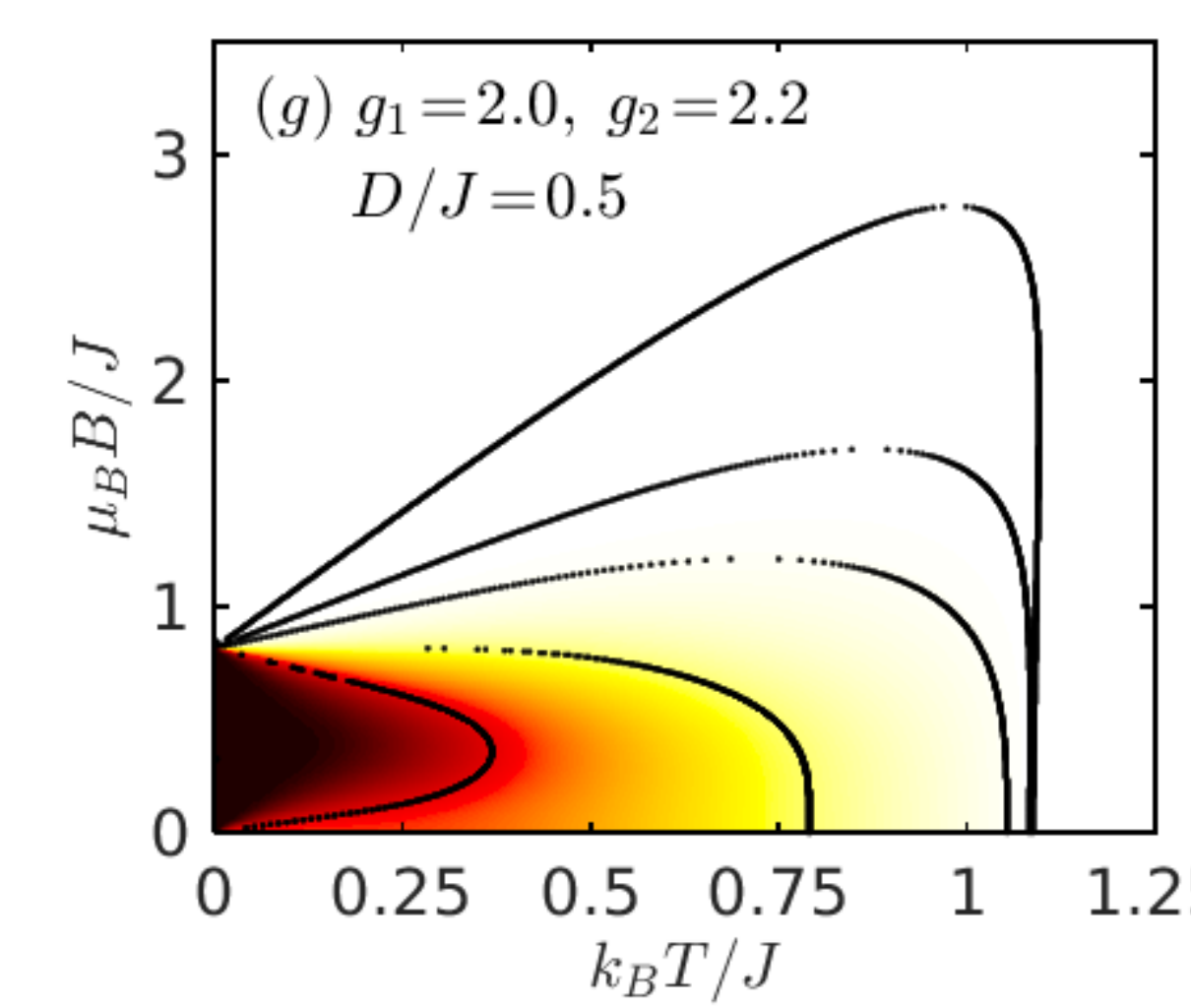}}
{\includegraphics[width=.288\textwidth,trim=3.2cm 9.8cm 2.8cm 9.5cm, clip]{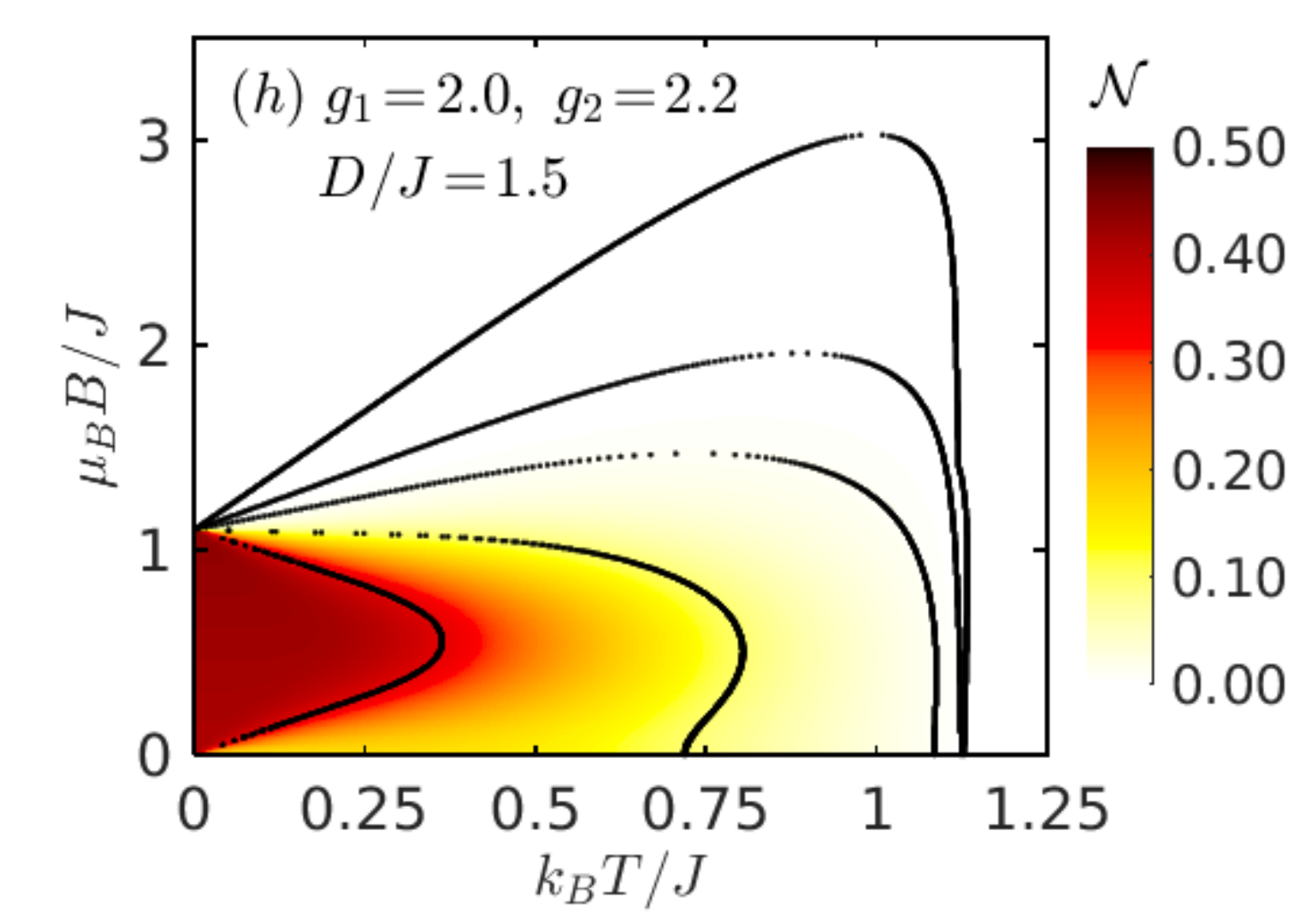}}
\caption{Density plots of the negativity ${\cal N}$ in the $k_BT/J$-$\mu_BB/J$ plane for $\Delta=1.0$, two different set of the Land\'e $g$-factors $g_1>g_2$ and $g_1<g_2$ with the same relative difference $|g_1-g_2|=0.2$ and several values of the uniaxial single-ion anisotropy $D/J$. The black contour lines correspond to ${\cal N}=$0.35, 0.1, 0.01, 0.001  and $10^{-5}$ (from left to right).}
\label{fig7}
\end{figure*}

To provide a deeper insight, density plots of the negativity are depicted in Fig.~\ref{fig7} for the fixed value of the exchange anisotropy $\Delta=1.0$, four different values of the uniaxial single-ion anisotropy $D/J$ and two different sets of Land\'e $g$-factors $g_1>g_2$ and $g_1<g_2$, respectively. In agreement with general expectations, the negativity mostly decreases upon increasing of temperature or magnetic field. The only exceptions to this rule apply to the magnetic fields slightly exceeding the saturation field when the thermal entanglement is enhanced upon increasing of temperature, as well as, to low enough temperatures when the uprise of magnetic field gives rise to an enhancement of the thermal entanglement. It could be thus concluded that the negativity shows qualitatively the same generic features for both settings of the $g$-factors as discussed previously for the particular case $g_1=g_2$. The marked difference in the respective density plots occurs just at relatively high temperatures $k_B T/J \gtrsim 1$ and magnetic fields $\mu_B B/J \gtrsim 1.5$, where a kink in contour lines of the negativity may be observed. Note furthermore that this non-trivial feature appears at very small values of the negativity ${\cal N} \lesssim 0.001$ just for $D/J<1/2$ on assumption that $g_1>g_2$ [see Fig.~\ref{fig7}(a)-(c)], while the same anomaly of the negativity can be detected for $D/J>1/2$ only if $g_1<g_2$ [see Fig.~\ref{fig7}(h)]. However, it is questionable if such small value of the negativity is capable of the experimental detection. Last but not least, the contour plots shown in Fig.~\ref{fig7}(d) and (h) convincingly evidence that the negativity may be reinforced upon rising of the magnetic field also at relatively high temperatures  (e.g. $k_B T/ J \lesssim 1$ for $D/J=1.5$) whenever a sufficiently strong easy-plane single-ion anisotropy is considered.  

\subsection{Thermal entanglement in the CuNi complex}
In this part we will put forward a theoretical prediction for a degree of quantum and thermal entanglement of the heterodinuclear complex CuNi, which affords an appropriate experimental realization of the mixed spin-(1/2,1) Heisenberg dimer \cite{hagi99}. It has been verified in the previous studies that the magnetic properties of the CuNi complex can be faithfully reproduced by the mixed spin-(1/2,1) Heisenberg dimer with the relatively strong isotropic exchange constant $J/k_B$ = 141~K, the gyromagnetic g-factors $g_1 = 2.20$ for Cu$^{2+}$ and $g_2 = 2.29$ for Ni$^{2+}$ magnetic ions, respectively, while any clear signatures of the exchange ($\Delta = 1$) or uniaxial single-ion ($D/k_B = 0)$ anisotropy has not been found. In the following we will therefore adapt this set of the model parameters in order to make the relevant theoretical prediction for the bipartite entanglement of the CuNi dimeric compound.
\begin{figure}[h!]
{\includegraphics[width=.9\columnwidth,trim=3cm 9.3cm 3.5cm 9.5cm, clip]{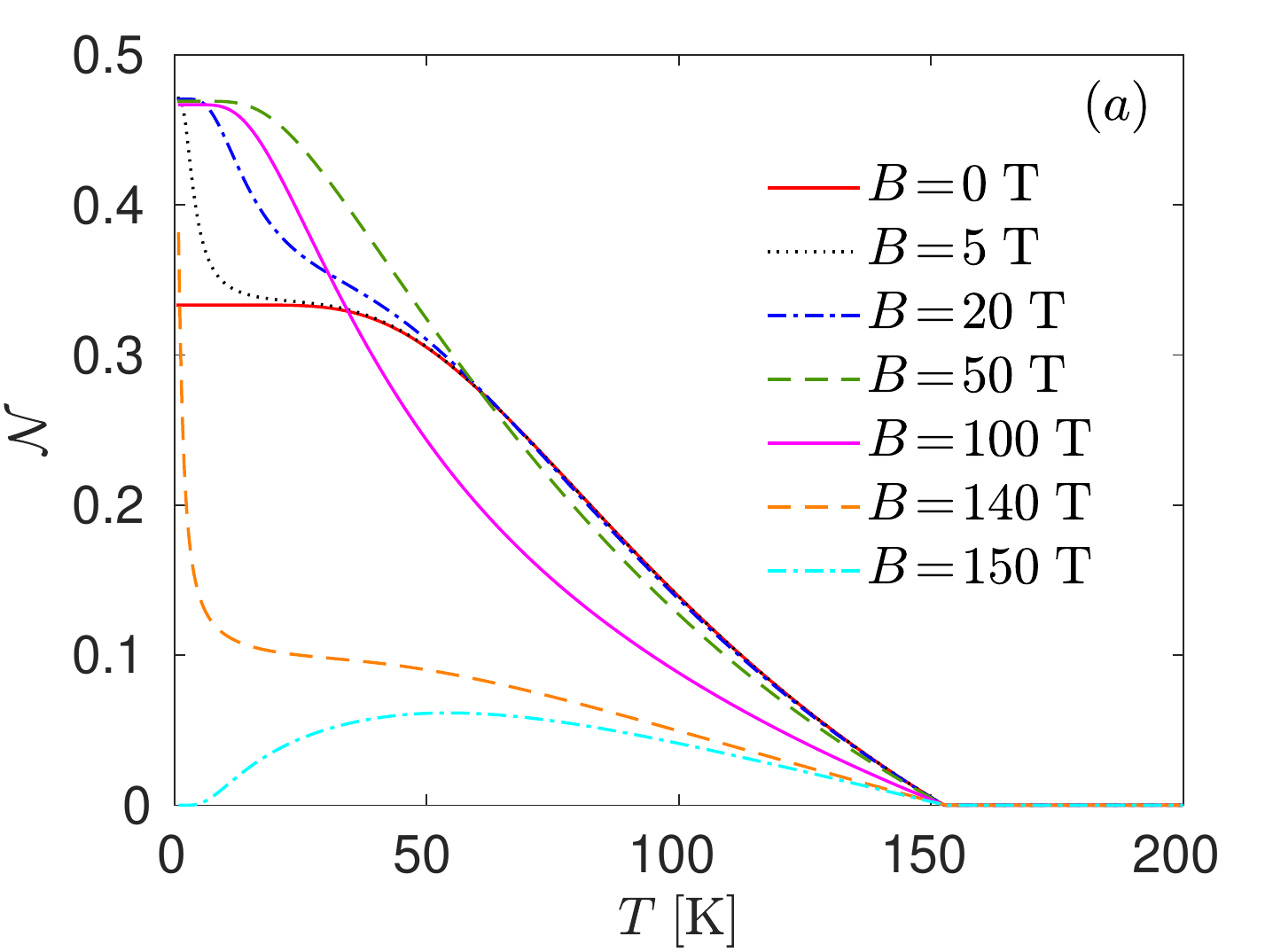}}
{\includegraphics[width=.9\columnwidth,trim=3cm 9.3cm 3.5cm 9.5cm, clip]{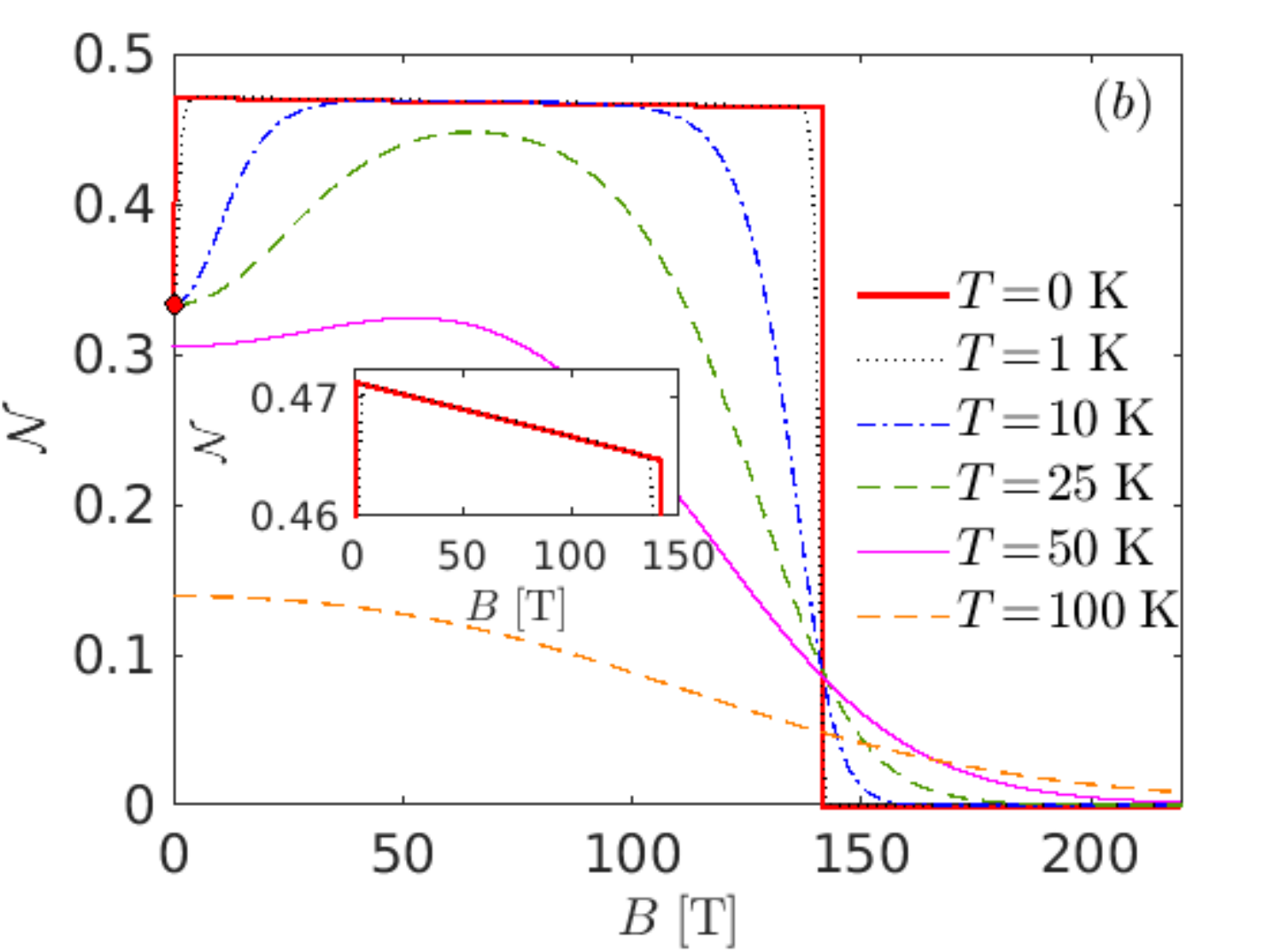}}
\vspace*{-0.3cm}
\caption{(a) Temperature dependences of the negativity ${\cal N}$ conforming to the CuNi compound for several values of the magnetic field; (b) magnetic-field dependences of the negativity ${\cal N}$ conforming to the CuNi compound for several values of temperature (the inset shows in an enhanced scale a quasi-plateau region). All displayed dependences were obtained for the mixed spin-(1/2,1) Heisenberg dimer with the isotropic exchange constant $J/k_B$ = 141~K ($\Delta = 1$), a zero single-ion anisotropy ($D/k_B$ = 0~K), the gyromagnetic g-factors $g_1 = 2.20$ and $g_2 = 2.29$ adapted according to Ref. \cite{hagi99}.}
\label{fig8}
\end{figure}

The negativity of the mixed spin-(1/2,1) Heisenberg dimer with the isotropic exchange constant $J/k_B$ = 141~K, the gyromagnetic g-factors $g_1 = 2.20$ and $g_2 = 2.29$ is plotted in Fig. \ref{fig8} as a function of temperature for a few selected values of the magnetic field and, respectively, as a function of the magnetic field for a few selected temperatures. Temperature variations of the negativity displayed in Fig. \ref{fig8}(a) exhibit mostly a monotonous decline with increasing temperature. At zero magnetic field the negativity monotonically decreases from the initial value ${\cal N} = 1/3$ until it completely vanishes at the threshold temperature $k_B T_t/J \approx 150$~K. At nonzero magnetic fields the negativity markedly starts from almost maximal value ${\cal N} \approx 0.47$, whereas the threshold temperature turns out to be independent of the magnetic field. An outstanding nonmonotonous thermal dependence of the negativity can be found only if the magnetic field surpasses the saturation value. Under this condition, the negativity become nonzero just at a lower threshold temperature, then it rises to a local maximum, which is successively followed by a gradual reduction until it repeatedly disappears at an upper threshold temperature (see the curve for $B = 150$~T). 

The isothermal dependence of the negativity on a magnetic field shown in Fig. \ref{fig8}(b) corroborates a transient strengthening of the thermal entanglement due to the external magnetic field. Owing to a difference of the gyromagnetic g-factors $g_1 = 2.20$ and $g_2 = 2.29$ of Cu$^{2+}$ and Ni$^{2+}$ magnetic ions, the negativity exhibits at very low temperatures $T \lesssim 1$~K a quasi-linear decrease (quasi-plateau) [see the inset in Fig. \ref{fig8}(b)], which is quite analogous to a quasi-plateau predicted for low-temperature magnetization curves of quantum Heisenberg spin systems with different g-factors \cite{ohan15}. Moreover, the negativity starts at sufficiently low temperatures $T \lesssim 25$~K from the initial value ${\cal N} = 1/3$, then it gradually increases to its local maximum before it finally diminishes upon further increase of the magnetic field. It is noteworthy that the initial value of the negativity is suppressed and its local maximum becomes more flat at moderate temperatures (e.g. for $T = 50$~K), while the negativity monotonically decreases upon strengthening of the external magnetic field at higher temperatures (e.g. for $T = 100$~K).   

\begin{figure}[h!]
{\includegraphics[width=.45\textwidth,trim=3.cm 9.5cm 3.cm 9.5cm, clip]{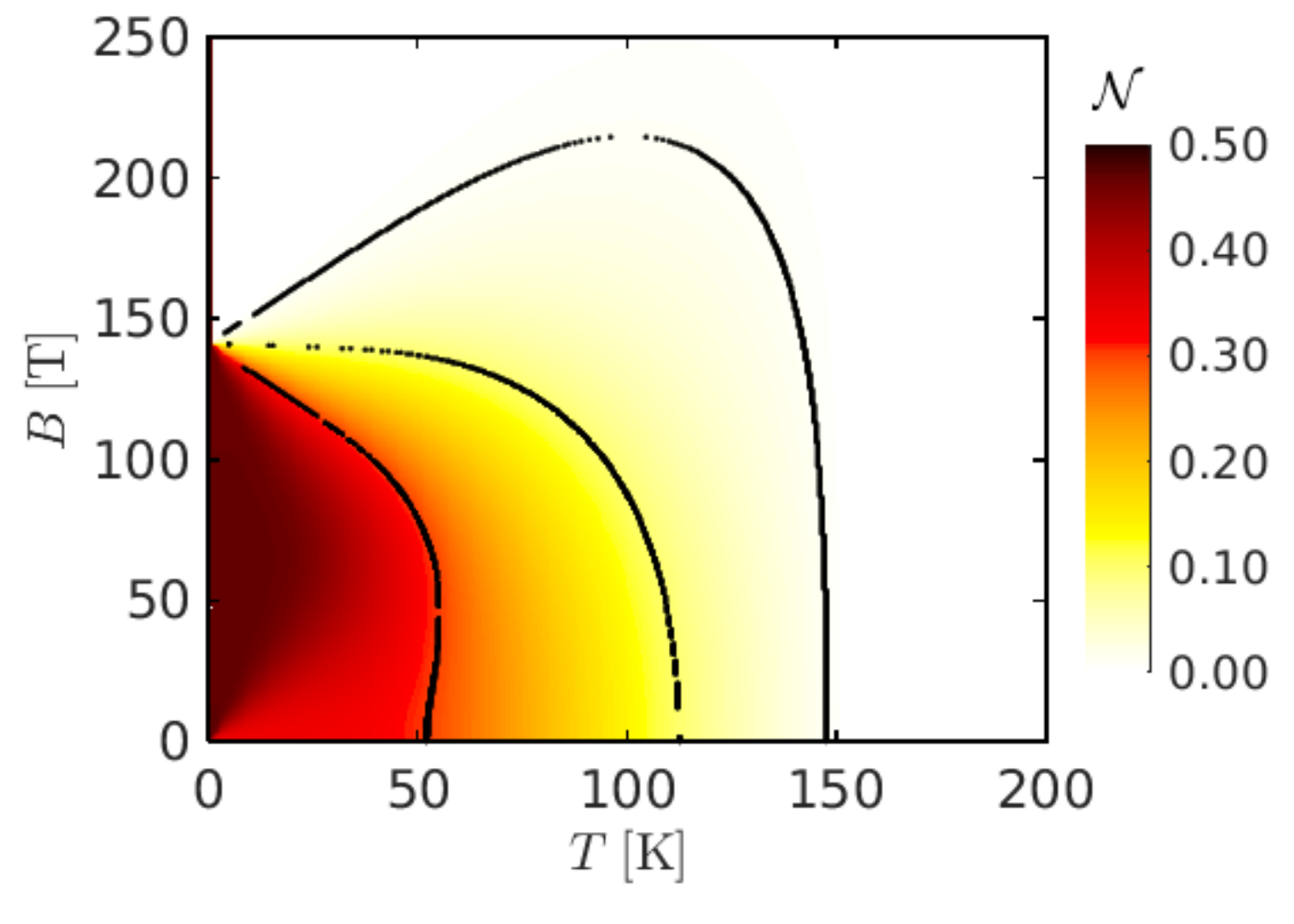}}
\vspace*{-0.3cm}
\caption{A density plot of the negativity ${\cal N}$ in the temperature-field plane conforming to the CuNi compound. The presented plot was obtained for the mixed spin-(1/2,1) Heisenberg dimer with the isotropic exchange constant $J/k_B$ = 141~K ($\Delta = 1$), a zero single-ion anisotropy ($D/k_B$ = 0~K), the gyromagnetic g-factors $g_1 = 2.20$ and $g_2 = 2.29$ adapted according to Ref. \cite{hagi99}. Black contour lines correspond to the particular values: ${\cal N}\!=$ 0.3, 0.1 and 0.01 (from left to right).}
\label{fig9}
\end{figure}
Last but not least, the density plot of the negativity in the temperature-field plane is depicted in Fig. \ref{fig9} for the mixed spin-(1/2,1) Heisenberg dimer with the isotropic exchange constant $J/k_B$ = 141~K ($\Delta = 1$), the gyromagnetic g-factors $g_1 = 2.20$ and $g_2 = 2.29$, which correspond according to Ref. \cite{hagi99} to the heterodinuclear complex CuNi. The displayed density plot can be alternatively viewed as a kind of 'phase diagram', which circumscribes a parameter space with a nonzero thermal entanglement from a disentangled parameter region. Although a subtle thermal entanglement can be detected even under extremely high magnetic fields and temperatures, the indispensable thermal entanglement of sufficient intensity (say ${\cal N} \gtrsim 0.1$) is confined to the magnetic fields $B < 140$~T and temperatures $T < 140$~K being comparable with the relevant exchange constant. While the bound set for the magnetic field considerably exceeds a reasonable range of magnetic fields for possible technological applications, the respective bound set for temperature apparently indicates necessity search for heterodinuclear complexes quite analogous to the CuNi compound \cite{hagi99}, which would however possess at least twice as large exchange constant in order to make technological applications at room temperatures vital.        

\section{Conclusion}
\label{sec:conc}

In the present article we have exactly examined the negativity of a mixed spin-(1/2,1) Heisenberg dimer, which quantifies a strength of the bipartite quantum and thermal entanglement at zero as well as nonzero temperatures within pure and mixed states of this simple quantum spin system. It has been evidenced that the negativity basically depends on intrinsic parameters as for instance exchange and uniaxial single-ion anisotropy in addition to extrinsic parameters such as temperature and magnetic field. The strongest quantum entanglement at zero temperature and zero magnetic field has been found for the particular case without uniaxial single-ion anisotropy and a perfectly isotropic coupling constant, while the negativity becomes completely independent of the exchange anisotropy for the specific strength of the uniaxial single-ion anisotropy $D/J = 1/2$. In presence of the external magnetic field the situation becomes much more intricate, because the negativity depends on gyromagnetic g-factors in addition to the exchange and uniaxial single-ion anisotropy, magnetic field and temperature. It turns out that the particular case with equal Land\'e g-factors exhibits the maximal quantum entanglement whenever the uniaxial single-ion anisotropy acquires the value $D/J = 1/2$.

In contrast to general expectations the rising magnetic field remarkably reinforces the bipartite quantum entanglement due to the Zeeman splitting of energy levels, which lifts a two-fold degeneracy of the quantum ferrimagnetic ground state. The maximal quantum entanglement is thus reached within a quantum ferrimagnetic phase at sufficiently low but nonzero magnetic fields on assumption that the gyromagnetic g-factors are equal and the uniaxial single-ion anisotropy is a half of the exchange constant $D/J = 1/2$. A strength of the bipartite quantum entanglement for the particular case with unequal gyromagnetic g-factors shows a quasi-linear dependence on the external magnetic field, which is quite reminiscent of a quasi-plateau phenomenon reported previously for low-temperature magnetization curves of quantum spin systems being composed of entities with unequal gyromagnetic g-factors \cite{ohan15}. It should be pointed out that all aforementioned generic trends are preserved for the bipartite thermal entanglement within the mixed states emergent at finite temperatures.  

The heterodinuclear complex CuNi as a prominent experimental representative of the mixed spin-(1/2,1) Heisenberg dimer afforded useful playground for an investigation of the bipartite thermal entanglement in a real-world system. It appears worthwhile to remark that the dimeric complex CuNi remains strongly entangled up to relatively high temperatures (about 140~K) and high magnetic fields (about 140~T) being comparable with the relevant exchange constant. From this point of view, the magnitude of the coupling constant in the heterodinuclear complex CuNi is inadequate for prospective technological applications of this solid-state material in quantum computing and quantum information processing at a room temperature. An enhancement of the coupling constant through a targeted design of some related heterodinuclear coordination compound of the type CuNi thus represents challenging task for material scientists.  

\begin{acknowledgements}
This work was financially supported by the grant of the Slovak Research and Development Agency provided under the contract No. APVV-18-0197 and by the grant of The Ministry of Education, Science, Research, and Sport of the Slovak Republic provided under the contract No. VEGA 1/0105/20.
\end{acknowledgements}

\begin{widetext}
\appendix
\section{}
\label{App A}
\setcounter{equation}{0}
\renewcommand{\theequation}{A\thesection.\arabic{equation}}
The explicit form of non-zero elements $\rho_{ij}$ of the density matrix given by Eq.~(\ref{eq12}). 
\allowdisplaybreaks
\begin{align}
\rho_{11}&=\frac{1}{Z}{\rm e}^{-\frac{\beta}{2} \left[J+2D-\left(h_{1}+2h_2\right)\right]};
\label{eqA1}\\
\rho_{22}&=\frac{1}{Z}{\rm e}^{\frac{\beta}{4} \left(J-2D+2h_{2}\right)}\left[ \cosh\left(\frac{\beta}{4}\sqrt{\left[J\!-\!2D\!-\!2(h_1\!-\!h_2)\right]^2\!+\!8(J\Delta)^2}\right)
\!-\!\frac{J\!-\!2D\!-\!2(h_1\!-\!h_2)}{\sqrt{\left[J\!-\!2D\!-\!2(h_1\!-\!h_2)\right]^2\!+\!8(J\Delta)^2}}\right.
\nonumber\\
&\times\left.\sinh\left(\frac{\beta}{4}\sqrt{\left[J\!-\!2D\!-\!2(h_1\!-\!h_2)\right]^2\!+\!8(J\Delta)^2}\right) \right];
\label{eqA2}\\
\rho_{33}&=\frac{1}{Z}{\rm e}^{\frac{\beta}{4} \left(J-2D-2h_{2}\right)}\left[ \cosh\left(\frac{\beta}{4}\sqrt{\left[J\!-\!2D\!+\!2(h_1\!-\!h_2)\right]^2\!+\!8(J\Delta)^2}\right)
\!+\!\frac{J\!-\!2D\!+\!2(h_1\!-\!h_2)}{\sqrt{\left[J\!-\!2D\!+\!2(h_1\!-\!h_2)\right]^2\!+\!8(J\Delta)^2}}\right.
\nonumber\\
&\times\left.\sinh\left(\frac{\beta}{4}\sqrt{\left[J\!-\!2D\!+\!2(h_1\!-\!h_2)\right]^2\!+\!8(J\Delta)^2}\right) \right];
\label{eqA4}\\
\rho_{44}&=\frac{1}{Z}{\rm e}^{\frac{\beta}{4}\left(J-2D+2h_{2}\right)}\left[ \cosh\left(\frac{\beta}{4}\sqrt{\left[J\!-\!2D\!-\!2(h_1\!-\!h_2)\right]^2\!+\!8(J\Delta)^2}\right)
\!+\!\frac{J\!-\!2D\!-\!2(h_1\!-\!h_2)}{\sqrt{\left[J\!-\!2D\!-\!2(h_1\!-\!h_2)\right]^2\!+\!8(J\Delta)^2}}\right.
\nonumber\\
&\times\left.\sinh\left(\frac{\beta}{4}\sqrt{\left[J\!-\!2D\!-\!2(h_1\!-\!h_2)\right]^2\!+\!8(J\Delta)^2}\right) \right];
\label{eqA6}\\
\rho_{55}&=\frac{1}{Z}{\rm e}^{\frac{\beta}{4} \left(J-2D-2h_{2}\right)}\left[ \cosh\left(\frac{\beta}{4}\sqrt{\left[J\!-\!2D\!+\!2(h_1\!-\!h_2)\right]^2\!+\!8(J\Delta)^2}\right)
\!-\!\frac{J\!-\!2D\!+\!2(h_1\!-\!h_2)}{\sqrt{\left[J\!-\!2D\!+\!2(h_1\!-\!h_2)\right]^2\!+\!8(J\Delta)^2}}\right.
\nonumber\\
&\times\left.\sinh\left(\frac{\beta}{4}\sqrt{\left[J\!-\!2D\!+\!2(h_1\!-\!h_2)\right]^2\!+\!8(J\Delta)^2}\right) \right];
\label{eqA7}\\
\rho_{66}&=\frac{1}{Z}{\rm e}^{-\frac{\beta}{2} \left[J+2D+\left(h_{1}+2h_2\right)\right]};
\label{eqA8}\\
\rho_{24}&=\rho_{42}\!=\!-\frac{\sqrt{8}J\Delta {\rm e}^{\frac{\beta}{4} \left(J-2D+2h_{2}\right)}}{Z\sqrt{\left[J\!-\!2D\!-\!2(h_1\!-\!h_2)\right]^2\!+\!8(J\Delta)^2}}\sinh\left(\frac{\beta}{4}\sqrt{\left[J\!-\!2D\!-\!2(h_1\!-\!h_2)\right]^2\!+\!8(J\Delta)^2}\right);
\label{eqA3}\\
\rho_{35}&=\rho_{53}\!=\!-\frac{\sqrt{8}J\Delta {\rm e}^{\frac{\beta}{4} \left(J-2D-2h_{2}\right)}}{Z\sqrt{\left[J\!-\!2D\!+\!2(h_1\!-\!h_2)\right]^2\!+\!8(J\Delta)^2}}\sinh\left(\frac{\beta}{4}\sqrt{\left[J\!-\!2D\!+\!2(h_1\!-\!h_2)\right]^2\!+\!8(J\Delta)^2}\right).
\label{eqA5}
\end{align} 
\end{widetext}

\end{document}